\documentclass[fleqn,usenatbib,useAMS]{mnras}

\DeclareRobustCommand{\VAN}[3]{#2}
\let\VANthebibliography\thebibliography
\def\thebibliography{\DeclareRobustCommand{\VAN}[3]{##3}\VANthebibliography}

\usepackage{amsmath,amsfonts,amssymb} 
\usepackage{bm} 
\usepackage[table,  dvipsnames]{xcolor} 
\usepackage{hyperref} 
\usepackage{nicefrac} 
\usepackage{placeins}

\usepackage[varg]{txfonts} 
\usepackage[T1]{fontenc}

\hypersetup{
	pdftitle = {Buoyancy torques prevent low-mass planets from stalling in low-turbulence radiative disks},
	pdfauthor = {Alexandros Ziampras, Sijme-Jan Paardekooper, Richard P. Nelson},
	pdfsubject = {},
	colorlinks = {true},
	linkcolor = [rgb]{0,0.35,0.7},
	citecolor = [rgb]{0,0.35,0.7},
	filecolor = [rgb]{0.61,0,0},
	urlcolor = [rgb]{0.61,0,0},
}

\usepackage{graphicx}  
\graphicspath{{figures/}}

\newcommand{\DP}[2]{\frac{\partial{#1}}{\partial{#2}}}
\newcommand{\DPP}[2]{\frac{\partial^2{#1}}{\partial{#2}^2}}
\newcommand{\D}[2]{\frac{\text{d}{#1}}{\text{d}{#2}}}

\newcommand{\G}{\text{G}}
\newcommand{\Mstar}{M_\star}
\newcommand{\Lstar}{L_\star}
\newcommand{\Rp}{R_\mathrm{p}}
\newcommand{\Mp}{M_\mathrm{p}}
\newcommand{\hp}{h_\mathrm{p}}
\newcommand{\Hp}{H_\mathrm{p}}

\newcommand{\Mth}{M_\mathrm{th}}
\newcommand{\Msun}{\mathrm{M}_\odot}
\newcommand{\Lsun}{\mathrm{L}_\odot}

\newcommand{\Rgas}{\mathcal{R}}
\newcommand{\cs}{c_\mathrm{s}}
\newcommand{\csiso}{c_\mathrm{s,iso}}

\newcommand{\OmegaK}{\Omega_\mathrm{K}}

\newcommand{\kappaR}{\kappa_\mathrm{R}}
\newcommand{\kappaP}{\kappa_\mathrm{P}}
\newcommand{\cv}{c_\mathrm{v}}
\newcommand{\rhomid}{\rho_\mathrm{mid}}
\newcommand{\sigmaSB}{\sigma_\mathrm{SB}}
\newcommand{\vel}{\bm{u}}
\newcommand{\xh}{{x}_\mathrm{h}}

\newcommand{\tcool}{t_\mathrm{cool}}
\newcommand{\bcool}{\beta_\mathrm{cool}}

\newcommand{\btot}{\beta_\mathrm{tot}}
\newcommand{\bcoll}{\beta_\mathrm{coll}}
\newcommand{\bbuoy}{\beta_\mathrm{buoy}}
\newcommand{\bdiff}{\beta_\text{diff}}

\newcommand{\Qrad}{Q_\mathrm{rad}}
\newcommand{\Qrelax}{Q_\mathrm{relax}}
\newcommand{\Erad}{E_\mathrm{rad}}

\newcommand{\aR}{a_\mathrm{R}}
\newcommand{\lrad}{l_\mathrm{rad}}

\newcommand{\omegas}{\omega_\mathrm{s}}
\newcommand{\omegaR}{\omega_\mathrm{R}}
\newcommand{\omegaP}{\omega_\mathrm{P}}
\newcommand{\omegac}{\omega_\mathrm{c}}

\newcommand{\ad}{a_\mathrm{d}}
\newcommand{\rhod}{\rho_\mathrm{d}}
\newcommand{\sd}{s_\mathrm{d}}
\newcommand{\md}{m_\mathrm{d}}
\newcommand{\pluto}{\texttt{PLUTO}}
\newcommand{\fargo}{{\texttt{FARGO3D}}}

\title[Buoyancy torques in radiative disks]{Buoyancy torques prevent low-mass planets from stalling in low-turbulence radiative disks}

\author[A.~Ziampras et al.]{
Alexandros~Ziampras$^{1}$\thanks{E-mail: a.ziampras@qmul.ac.uk},
Richard~P.~Nelson$^{1}$,
Sijme-Jan~Paardekooper$^{1,2}$
\\
$^{1}$Astronomy Unit, School of Physics and Astronomy, Queen Mary University of London, London E1 4NS, UK\\
$^{2}$Faculty of Aerospace Engineering, Delft University of Technology, Kluyverweg 1, 2600 AA Delft, The Netherlands\\
}

\date{Accepted XXX. Received YYY; in original form ZZZ}
\pubyear{2023}

\begin{document}
\label{firstpage}
\pagerange{\pageref{firstpage}--\pageref{lastpage}}
\maketitle

\begin{abstract}
	Low-mass planets migrating inwards in laminar protoplanetary disks (PPDs) experience a dynamical corotation torque, which is expected to slow down migration to a stall. However, baroclinic effects can reduce or even reverse this effect, leading to rapid inward migration. In the radiatively inefficient inner disk, one such mechanism is the buoyancy response of the disk to an embedded planet. Recent work has suggested that radiative cooling can quench this response, but for parameters that are not necessarily representative of the inner regions of PPDs. We perform global three dimensional inviscid radiation hydrodynamics simulations of planet--disk interaction to investigate the effect of radiative cooling on the buoyancy-driven torque in a more realistic disk model. We find that the buoyancy response exerts a negative dynamical corotation torque --- albeit partially damped due to radiative cooling --- resulting in sustained, rapid inward migration. Models that adopt a local cooling prescription significantly overestimate the impact of the buoyancy response, highlighting the importance of a realistic treatment of radiation transport that includes radiative diffusion. Our results suggest that low-mass planets should migrate inwards faster than has been previously expected in radiative disks, with implications for the formation and orbital distribution of super-Earths and sub-Neptunes at intermediate distances from their host stars, unless additional physical processes that can slow down migration are considered.
\end{abstract}

\begin{keywords}
    planet--disc interactions --- protoplanetary discs --- hydrodynamics --- radiation: dynamics --- methods: numerical
\end{keywords}


\section{Introduction}
\label{sec:introduction}

Planets are born in protoplanetary disks (PPDs), and their observed location after the disk has dispersed is highly sensitive to their interaction with it. In particular, for low-mass planets, addressing their origin is a long-standing problem given how rapidly they migrate through the disk \citep{ward-1997a}. A solution to this problem is crucial in order to explain the large population of super-Earths and sub-Neptunes in the observed exoplanet demographics. For a recent review, see \citet{paardekooper-etal-2022}.

Regardless of planet mass, different mechanisms have been proposed to halt or slow the inward migration of young planets or even reverse it. Massive enough planets that can open a deep gap around their orbit \citep[type-II regime,][]{lin-papaloizou-1986,rafikov-2002} can migrate very slowly in disks that sustain very low levels of turbulent viscosity \citep{ward-1997a,lega-etal-2022} although interaction with the Rossby-wave unstable gap edge can lead to vortex-assisted inward or outward migration \citep{mcnally-etal-2019a,lega-etal-2022}. For intermediate-mass planets, migration has previously been expected to eventually stall as the planet acts akin to a snow-plough and shovels a wall of gas ahead of its corotating region, in a phenomenon dubbed the ``inertial limit'' \citep{hourigan-ward-1984,ward-hourigan-1989,ward-1997a}. However, more recent work has shown that vortex activity can help sustain inward migration \citep{mcnally-etal-2019a}.

For low-mass planets (type-I regime), the disk remains relatively unperturbed by the planet's presence beyond the formation of spiral arms \citep{ogilvie-lubow-2002} and the planet's corotating region \citep{goldreich-tremaine-1980}. Here, corotation torques have been shown to efficiently stop or even reverse migration when the turbulent viscosity is sufficiently high \citep{paardekooper-etal-2011}, and thermal torques due to the planet's accretion luminosity can do the same for roughly Earth-mass planets \citep{benitez-etal-2015,masset-2017}. For super-Earths and sub-Neptunes in laminar disks however, none of these effects operate efficiently enough to significantly slow down type-I migration.

In laminar disks, where accretion is driven through the disk surface in the form of a magnetothermal wind \citep{bai-stone-2013,gressel-2015,wang-etal-2019}, the corotation torque described above vanishes and the planet is forced to drag its corotating region along as it migrates inwards \citep{paardekooper-2014}. As long as the vortensity of the material enclosed in that region is conserved (i.e., in the absence of baroclinic effects), the associated drag takes the form of a dynamical corotation torque (DCT) and can slow the planet down to an effective halt \citep{paardekooper-2014} even though migration doesn't formally stop.

This last scenario could help reconcile theoretical modeling with exoplanet demographics by dramatically slowing down type-I migration, thereby preventing low-mass planets from migrating to the inner rim of the disk. However, as stated above, it relies on the conservation of vortensity in the corotating region. While this condition is satisfied for barotropic flows (e.g., isothermal disks), radiative effects have been shown to generate vortensity near the planet \citep{pierens-2015,ziampras-etal-2024}, resulting in a negative torque that can not only reduce the drag due to the DCT but even accelerate inward migration beyond the expected type-I speed. At the same time, the buoyancy response of the disk to a low-mass embedded planet has been shown to produce a similar vorticity-generating effect in radiatively inefficient, adiabatic disks \citep{mcnally-etal-2020}, in addition to providing an apparently smaller and separate buoyancy torque that arises through direct interaction between the planet and the buoyant modes in the disk \citep{zhu-etal-2012,lubow-zhu-2014}. Recent work has shown that, in the context of a specific disk model, radiative cooling can act to quench this buoyancy response, preventing vortensity generation and erasing its contribution to the total torque \citep{yun-etal-2022}.

While hydrodynamical modeling with a treatment of radiation transport is far more realistic than a simplified isothermal or adiabatic equation of state, it is significantly more expensive and the results are highly sensitive to the underlying disk model, which determines the cooling timescale and therefore the efficiency of radiative processes. With that in mind, while \citet{yun-etal-2022} have demonstrated that radiative cooling can damp the disk buoyancy response, and given that its operation can decide the fate of embedded, low-mass planets \citep{paardekooper-2014}, it is crucial to investigate in which disk conditions their result holds true. As shown by \citet{yun-etal-2022}, this requires simulations that treat radiative cooling in a realistic way.

Therefore, in this study, we perform hydrodynamical simulations that feature the disk buoyancy response to an embedded planet, designed to represent the inner few au of the disk and with a realistic treatment of radiation transport. Our goal is to examine whether the buoyancy-associated torque is ever relevant in real disks, and, if so, to what extent it is capable of modifying the planet's migration rate.

We describe our physical framework in Sect.~\ref{sec:physics-numerics}. We motivate our numerical parameters in Sect.~\ref{sec:cooling-timescale-maps}, present our results from numerical simulations in Sect.~\ref{sec:results-simulations}, and discuss their implications in Sect.~\ref{sec:discussion}. We finally summarize our findings in Sect.~\ref{sec:summary}.

\section{Physical framework}
\label{sec:physics-numerics}

In this section we lay out the physical framework of our models. We describe in detail our approach to radiation transport, and motivate our numerical experiments by investigating the conditions under which buoyancy torques could be relevant.

\subsection{Equations of hydrodynamics}
\label{sub:equations-of-hydrodynamics}

We consider a disk of ideal gas with adiabatic index $\gamma=\nicefrac{7}{5}$ and mean molecular weight $\mu=2.353$, orbiting a star with mass $\Mstar=\Msun$ and luminosity $\Lstar=\Lsun$. The volume density $\rho$, velocity field $\vel$, and pressure $P$ of the gas evolve according to the (inviscid) Euler equations
\begin{subequations}
	\label{eq:navier-stokes}
	\begin{align}
		\label{eq:navier-stokes-1}
		\DP{\rho}{t} + \vel\cdot\nabla\rho=-\rho\nabla\cdot\vel,
	\end{align}
	\begin{align}
	\label{eq:navier-stokes-2}
		\DP{\vel}{t}+ (\vel\cdot\nabla)\vel=-\frac{1}{\rho}\nabla P -\nabla\Phi,
	\end{align}
	\begin{align}
	\label{eq:navier-stokes-3}
		\DP{e}{t} + \vel\cdot\nabla e=-\gamma e\nabla\cdot\vel + Q,
	\end{align}
\end{subequations}
where $e=P/(\gamma-1)$ is the internal energy density given by the ideal gas law, $\Phi = \Phi_\star = \G \Mstar/r$ is the gravitational potential of the star at distance $r$, $\G$ is the gravitational constant, and $Q$ encapsulates any additional sources of heating or cooling. The isothermal sound speed is then $\csiso=\sqrt{P/\rho}$, and the temperature is $T=\mu \csiso^2/\Rgas$, with $\Rgas$ denoting the gas constant.

In a cylindrical coordinate system $(R,\varphi,z)$ with $r=\sqrt{R^2+z^2}$, by assuming that the disk is non-accreting and axisymmetric and further requiring that the midplane density $\rhomid$ and vertically constant temperature follow radial power-law profiles such that
\begin{equation}
	\label{eq:rho-T}
	\rhomid(R) = \rho_0 \left(\frac{R}{R_0}\right)^p, \quad T(R) = T_0 \left(\frac{R}{R_0}\right)^q,
\end{equation}
the disk structure in equilibrium can be described by \citep{nelson-etal-2013}
\begin{align}
	\label{eq:equilibrium}
	\rho^\text{eq}(R,z) &= \rhomid(R)\, \exp\left[-\frac{1}{h^2}\left(1-\frac{R}{r}\right)\right], \\
	u_\phi(R,z) &= R\OmegaK \left[1 + (p+q)h^2 + q\left(1-\frac{R}{r}\right)\right]^{1/2}.
\end{align}
Here, $\OmegaK=\sqrt{\G\Mstar/R^3}$ is the Keplerian angular velocity, $h=H/R$ is the aspect ratio of the disk, and $H=\csiso/\OmegaK$ is the pressure scale height. Finally, we can define the surface density through $\Sigma=\int_{-\infty}^\infty\rho\,\mathrm{d}z$.

\subsection{Radiative cooling}
\label{sub:radiative-cooling}

We model the radiative cooling of the disk using the flux-limited diffusion (FLD) approximation \citep{levermore-pomraning-1981}. In this approach, the radiation energy density $\Erad$ evolves as
\begin{equation}
	\label{eq:Erad}
	\DP{\Erad}{t} + \nabla\cdot\bm{F}=\kappaP\rho c\left(\aR T^4 - \Erad\right),
\end{equation}
and is coupled to the gas energy density with a source term
\begin{equation}
	\label{eq:Qrad}
	\Qrad = -\kappaP\rho c\left(\aR T^4 - \Erad\right)
\end{equation}
in Eq.~\eqref{eq:navier-stokes-3}. Here, $\kappaP$ is the Planck mean opacity, $c$ is the speed of light, and $\aR$ is the radiation constant. The radiation flux $\bm{F}$ is given by
\begin{equation}
	\label{eq:flux}
	\bm{F} = -\frac{\lambda c}{\kappaR\rho}\nabla\Erad,
\end{equation}
where $\kappaR$ is the Rosseland mean opacity and $\lambda$ is the flux limiter following \citet{kley-1989}:
\begin{equation}
	\label{eq:flux-limiter}
	\lambda = \begin{cases}
        \frac{2}{3+\sqrt{9+10 x^2}}, & x \leq 2, \\
        \frac{10}{10x + 9 + \sqrt{180x+81}}, & x > 2.
    \end{cases},\quad
	x := \frac{|\nabla\Erad|}{\kappaR\rho\Erad}.
\end{equation}

We do not explicitly include the heating due to stellar irradiation in our models. Although this implies that our disk model is only accurate up to the disk optical surface \citep[$z\sim4H$, see][]{chiang-goldreich-1997}, this is not a problem as we are mainly interested in the region $z\lesssim2H$. We can sidestep this issue by prescribing $\Erad=\aR T_0^4$ at the upper boundary of our disk which prevents the gas from rapidly cooling off (see also Sect.~\ref{sub:numerics}).

In the FLD approach, the cooling timescale $\bcool = \tcool\OmegaK$ can be approximated following \citet{flock-etal-2017b}
\begin{equation}
	\label{eq:beta}
	\bcool = \frac{\OmegaK}{\eta}\left(H^2 + \frac{\lrad^2}{3}\right),\qquad\eta=\frac{16\sigmaSB T^3}{3\kappaP\rho^2\cv},\quad \lrad = \frac{1}{\kappaP\rho}.
\end{equation}
Here, $\sigmaSB$ is the Stefan--Boltzmann constant, and $\cv=\Rgas/(\mu(\gamma-1))$ is the specific heat at constant volume.

The above expression for $\beta$ is valid under the assumption that the solid particles (``dust'') in the disk, which facilitate cooling, are well-coupled to the gas. In particular, we focus on the small grains with size $\ad=0.1\,\mathrm{\mu m}$, as they provide the most efficient cooling channel. We can then express the collision timescale $\bcoll$ between the gas and small grains following \citet{dullemond-etal-2022}
\begin{equation}
	\label{eq:bcoll}
	\bcoll = \frac{\cv e}{\epsilon\rho^2\bar{C}_H}\frac{\md}{\sd}\OmegaK, \qquad\bar{C}_H = \cv\frac{\csiso}{\sqrt{2\pi}}
\end{equation}
Here, $\sd=4\pi\ad^2$ and $\md=\frac{4\pi}{3}\rhod \ad^3$ are the surface area and mass of a single dust particle with bulk density $\rhod=2.08\,\mathrm{g}/\mathrm{cm}^3$, and $\epsilon$ is the dust-to-gas mass ratio. The combined cooling timescale $\btot$ is then given by
\begin{equation}
	\label{eq:btot}
	\btot = \bcool + \beta_\mathrm{coll}.
\end{equation}

We note that the cooling timescale is a sensitive function of the disk's temperature and density structure, and depends strongly on the choice of opacity model. In our models, we use the density- and temperature-dependent opacity model of \citet{lin-papaloizou-1985} with $\kappaR = \kappaP$.

\subsection{Buoyancy-driven torque}
\label{sub:buoyancy-driven-torque}
The gravitational interaction between disk and planet excites a buoyancy response in the disk, which can generate a torque on the planet. \citet{zhu-etal-2012} showed with local shearing box simulations that the instantaneous torque due to the direct interaction between the planet and the buoyantly excited gas can become comparable to the Lindblad torque \citep{ward-1997a} in the adiabatic limit ($\btot\to\infty$). \citet{mcnally-etal-2020} then carried out global adiabatic models with a migrating planet and showed that, in addition to the effect shown by \citet{zhu-etal-2012}, buoyancy oscillations result in a long-term build-up of vortensity ($\varpi$) in the planet's corotating region, which can erase or even reverse the stalling effect of the dynamical corotation torque \citep{paardekooper-2014,mcnally-etal-2017}
\begin{equation}
	\label{eq:DCT}
	\Gamma_h  = 2\pi\left(1-\frac{\varpi(\Rp)}{\varpi_h}\right)\Sigma_\mathrm{p}\Rp^2 \xh\Omega_\mathrm{p}\left(\D{\Rp}{t} - u_R\right).
\end{equation}
Here, a subscript ``p'' denotes quantities at the planet's location, $u_R$ is the radial velocity of the background disk, and $\xh$ is the horseshoe half-width \citep{paardekooper-etal-2010b} defined as
\begin{equation}
	\label{eq:horseshoe-width} 
	\xh = \frac{1.1}{\gamma^{1/4}}\left(\frac{0.4}{\epsilon/H_\text{p}}\right)^{1/4}\sqrt{\frac{\Mp}{\hp\Mstar}}\Rp,\qquad \epsilon = 0.6H_\text{p}.
\end{equation}
The smoothing length $\epsilon$ is chosen to match 3D models \citep{mueller-kley-2012}.

In Eq.~\eqref{eq:DCT}, the vortensity is given in the vertically integrated approximation by $\varpi=(\nabla\times \vel)/\Sigma\cdot\hat{z}$, while \citet{masset-llambay-2016} derived the equivalent quantity in 3D to be
\begin{equation}
	\label{eq:vortensity-3d}
	\varpi = \left[\int\limits_{-\infty}^\infty\frac{\rho}{(\nabla\times\vel)\cdot\hat{z}}\mathrm{d}z\right]^{-1}.
\end{equation}
Then, $\varpi(\Rp)$ is the vortensity at $\Rp$ for an unperturbed disk, and $\varpi_\mathrm{h}$ is the characteristic vortensity enclosed in the planet's corotating region. As the planet migrates inwards it carries with it the corotating material that was present at its initial location, while preserving the value of $\varpi_h$ in the absence of turbulent viscosity or vortensity-generating baroclinic effects. This in turns leads to evolution of the quantity ${\varpi(\Rp)}/{\varpi_h}$, and a change in the corotation torque.

In the absence of vortensity-generating (baroclinic) effects the unperturbed vortensity in the background disk is given by $\varpi_0(R) = \frac{1}{2}\OmegaK/\Sigma^\text{eq}$, which would result in a continuously increasing positive torque as the planet migrates inwards for $\Sigma(R)$ shallower than $R^{-3/2}$. However, the excitation of buoyancy oscillations in the planet's horseshoe region induces a baroclinic forcing \citep{ziampras-etal-2023b}, generates vortensity, and results eventually in a negative dynamical corotation torque \citep{mcnally-etal-2020}.

\citet{yun-etal-2022} have called into question the relevance of the buoyancy-driven torque in radiative disks, showing that radiative diffusion can quench the buoyancy response and erase the associated torque when the cooling timescale due to radiative diffusion ($\bdiff=\OmegaK H^2/\eta$, see Eq.~\eqref{eq:beta}) is comparable to or shorter than the buoyancy timescale $\bbuoy$. The latter is related to the Brunt--V\"ais\"al\"a frequency
\begin{equation}
	\label{eq:buoyancy-timescale}
	N = \sqrt{-\frac{1}{\gamma}\DP{\Phi_\star}{z}\DP{}{z}\left[\ln\left(\frac{P}{\rho^\gamma}\right)\right]} = \OmegaK \sqrt{\frac{\gamma-1}{\gamma}} \frac{z}{H} \left(\frac{R}{r}\right)^3,
\end{equation}
through which we can write
\begin{equation}
	\label{eq:bbuoy}
	\bbuoy = \frac{\OmegaK}{N} = \sqrt{\frac{\gamma}{\gamma-1}}\frac{H}{z}\left(1 + \frac{z}{R}\right)^{3/2}.
\end{equation}
In their simulations and for their disk model, \citet{yun-etal-2022} found that $\bdiff\lesssim\bbuoy$ for $z\gtrsim 2H$, and concluded that the buoyancy-driven torque is quenched in radiative disks. However, as stated above, $\bdiff$ is a sensitive function of the underlying disk model. It is therefore necessary to investigate the conditions under which the buoyancy-driven torque is relevant in a more slowly cooling, passively irradiated disk model \citep[e.g.,][]{chiang-goldreich-1997}.

\section{Cooling timescale maps}
\label{sec:cooling-timescale-maps}

In this section we present maps of the cooling timescale $\btot$ and the ratio $\bbuoy/\btot$ for several different disk models. We use these maps to motivate our choice of numerical parameters for our hydrodynamical simulations.

\subsection{Connecting to previous work}
\label{sub:colin-model}

\begin{figure*}
	\centering
	\includegraphics[width=\textwidth]{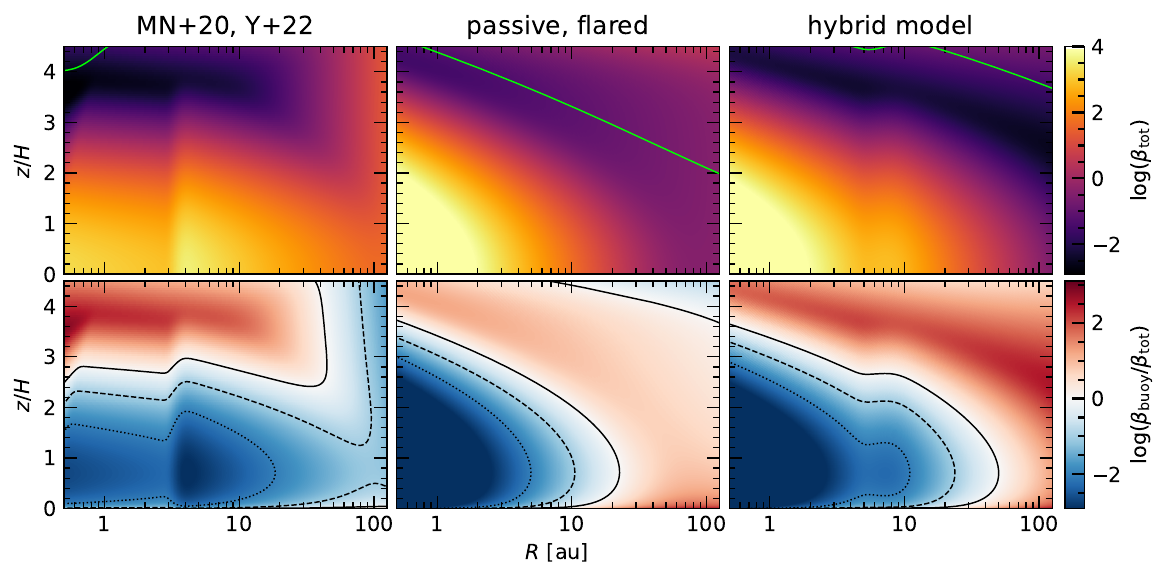}
	\caption{Cooling timescale $\btot$ (top) and the ratio to the buoyancy timescale $\bbuoy/\btot$ (bottom) for different disk models. Left: the model of \citet{mcnally-etal-2020} and \citet{yun-etal-2022}. Middle: a more realistic model of a passive, irradiated disk. Right: a hybrid model that combines the computational cost and reference parameters of model A with the radiative environment of model B. Green contours in the upper panels denote the ``optical surface'' of the disk, where $\tau_\mathrm{irr}(r)=\int_{R_\odot}^{r}\kappaP(T_\odot)\rho\,\mathrm{d}r=1$. Solid, dashed, and dotted black contours in the lower panels denote $\bbuoy/\btot=1$, 0.1, and 0.01, respectively.}
	\label{fig:disk-maps}
\end{figure*}

We first compute the cooling timescale $\btot$ for the disk model of \citet{mcnally-etal-2020} and \citet{yun-etal-2022}. Their model assumes a disk with $\Sigma(R) = 3400\,\left(R/\text{au}\right)^{-1/2}\,\text{g}/\text{cm}^2$, a constant aspect ratio $h=0.05$ (i.e., $T\propto1/R$) and a constant dust-to-gas mass ratio $\epsilon=0.01$ for submicron grains. Combining Eqs.~\eqref{eq:rho-T},~\eqref{eq:equilibrium},~\eqref{eq:beta}, and~\eqref{eq:bcoll} we compute the cooling timescale $\btot=\bcool+\bcoll$ and the ratio $\bbuoy/\btot$ for this disk model, and present the results in the left panels of Fig.~\ref{fig:disk-maps}.

In agreement with the findings of \citet{yun-etal-2022}, we find that $\bbuoy/\btot\sim0.1$ at $z\sim2H$ in the inner few au, which is why buoyancy torques were quenched in their disk model. In general, with the exception of a small region around the water iceline at 5\,au, we find that $\bbuoy/\btot\gtrsim0.01$ between $z\sim1\text{--}2H$ throughout the whole disk, and therefore expect that buoyancy torques are indeed not relevant for this disk model.

\subsection{A more realistic disk model}
\label{sub:realistic-disk-model}

The disk model used by \citet{mcnally-etal-2020}, while both simple and effective in demonstrating the effect of buoyancy torques in a scale-free environment, is not quite representative of a real PPD. Passively heated disks are flared \citep[$q\approx-1/2$,][]{chiang-goldreich-1997} and significantly thinner ($h\approx0.02\text{--}0.025$ at 1\,au) than the constant aspect ratio disk model used above. Achieving such a comparatively high value for $h$ would require considerable internal heating due to turbulence, with a corresponding $\alpha\sim10^{-3}\text{--}10^{-2}$ \citep{shakura-sunyaev-1973}. This is both difficult to reconcile with the low turbulence expected in the dead zone \citep{bai-stone-2013} and inconsistent with the assumption that the models by both \citet{mcnally-etal-2020} and \citet{yun-etal-2022} reflect laminar disks, where buoyancy torques can operate in the first place.

With that in mind, we repeat the above exercise assuming a passively heated disk model. We assume that the aspect ratio $h$ follows a power-law profile $h(R) = 0.02\,\left(R/\text{au}\right)^{2/7}$, or equivalently that $q=-3/7$ \citep{chiang-goldreich-1997,menou-goodman-2004}. We also assume that the surface density profile is given by $\Sigma(R) = 1700\,\left(R/\text{au}\right)^{-1}\,\text{g}/\text{cm}^2$, matching the Minimum Mass Solar Nebula model \citep{hayashi-1981} at $R=1$\,au. We then recompute the cooling timescale $\btot$ and the ratio $\bbuoy/\btot$, and present the results in the middle panels of Fig.~\ref{fig:disk-maps}.

Given that $\bdiff\propto T^{-3}\propto h^{-6}$, such a significant drop in the aspect ratio results in dramatically longer cooling timescales in the inner disk, even though the surface density is lower. In fact, we find that $\bbuoy/\btot\lesssim10^{-3}$ up to $z\sim2\text{--}3H$ for $R\lesssim2\,\text{au}$, which could be sufficient for buoyancy torques to operate in the inner disk.

\subsection{A computationally friendly hybrid model}
\label{sub:hybrid-model}

The model computed in the previous section, while more realistic and certainly more optimistic for the operation of buoyancy torques, is also significantly more expensive to use in the global models we would like to execute. It would also result in much more rapid migration as the Lindblad torque scales with $h^{-2}$ \citep{goldreich-tremaine-1979}. Given that the buoyancy response of the disk is a subtle effect highly sensitive to the numerical resolution \citep{ziampras-etal-2023b}, achieving a high enough resolution ($\sim$16 cells per scale height) to capture the effect of buoyancy torques in the inner disk ($h\sim0.02$) while also integrating the radiation hydrodynamics equations in Eqs.~\eqref{eq:navier-stokes}~\&~\eqref{eq:Erad} for several hundred planetary orbits is computationally prohibitive.

We note here that our goal is not to measure the precise migration rate of a planet in a realistic disk model, but rather to investigate whether the buoyancy-driven torque is even active in the inner few au of a disk, where there still exists the possibility that radiative diffusion can damp the buoyancy response of the disk. With the optimistic model in Sect.~\ref{sub:realistic-disk-model} in mind, we therefore choose to use a hybrid disk model that is both computationally friendly and still representative of the radiative environment in a real protoplanetary disk.

To that end, we assume a flared disk with $q=-3/7$ and $\Sigma\propto1/R$ as before, but adjust $\Sigma_0$ and $h_0$ such that the values of $\Sigma$ and $h$ at the planet's initial radial location of 2\,au are identical to those in the model of \citet{mcnally-etal-2020} and \citet{yun-etal-2022}. In this way, our model represents similar conditions to the realistic model in Sect.~\ref{sub:realistic-disk-model}, is computationally feasible, and also maintains compatibility with previous work to an extent. We therefore choose $\Sigma_0=4800~\text{g}/\text{cm}^2~(R/\text{au})^{-1}$ and $h=0.041~(R/\text{au})^{2/7}$.

\begin{figure}
	\centering
	\includegraphics[width=\columnwidth]{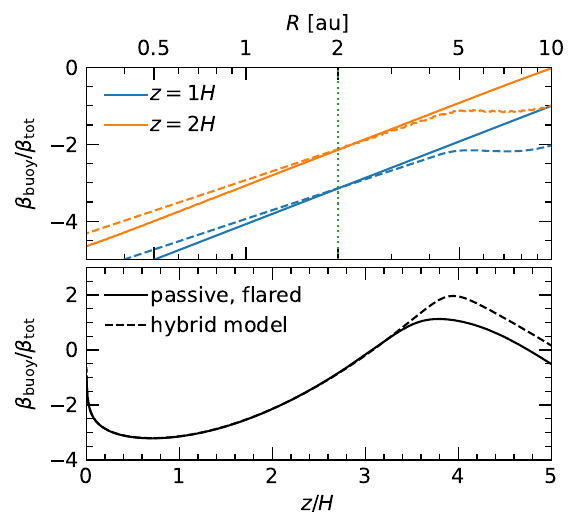}
	\caption{Comparison of the ratio $\bbuoy/\btot$ for the realistic and hybrid disk models in our region of interest, where buoyancy torques could operate ($R\lesssim4$\,au). Top: radial profiles of $\bbuoy/\btot$, showing a near-perfect match for $R\approx1$--2\,au. Bottom: vertical profile of $\bbuoy/\btot$ at $R=2$\,au.}
	\label{fig:hybrid-beta}
\end{figure}

Of course, adjusting both $\Sigma$ and $h$ in this way will inevitably also change the cooling timescale compared to our previous model. To account for this difference and ensure that the disk buoyancy response is similar between the model in Sect.~\ref{sub:realistic-disk-model} and this hybrid model, we artificially increase the dust opacity by a factor of $\approx7$.
The result is a near-perfect match between the two models in terms of the ratio $\bbuoy/\btot$ both radially and vertically, as shown in Fig.~\ref{fig:hybrid-beta}. We also show the cooling timescale maps for this model in the right panels of Fig.~\ref{fig:disk-maps}. We will be using this hybrid disk model in our hydrodynamical simulations in Sect.~\ref{sec:results-simulations}.

\section{Hydrodynamical simulations}
\label{sec:results-simulations}

In this section we present the results from our hydrodynamical simulations. We first describe our numerical setup, and then present the results from our models with a migrating planet.

\subsection{Numerical setup}
\label{sub:numerics}

We use the numerical hydrodynamics Godunov code \pluto{} \texttt{v.4.4} \citep{mignone-etal-2007} with a radiation transport module following the implementation of the FLD closure by \citet{kolb-etal-2013} and described in Appendix~\ref{apdx:FLD-test}. We also utilize the FARGO algorithm \citep{masset-2000}, implemented in \pluto{} by \citet{mignone-etal-2012}. This method alleviates the strict timestep limitation imposed by the rapidly rotating inner radial boundary by subtracting the Keplerian rotation profile of the background disk before advecting the gas. In doing so, it provides a substantial speedup while also reducing numerical diffusion.

We use the \texttt{ENTROPY\_SWITCH} option to offset the susceptibility of \pluto{} to the high-Mach problem \citep{trac-pen-2004}. We also use the HLLC Riemann solver \citep{harten-1983}, a second-order (\texttt{RK2}) time marching scheme, a third-order weighted essentially non-oscillatory reconstruction \citep[\texttt{WENO3},][]{yamaleev-carpenter-2009} and the flux limiter by \citet{vanleer-1974}. This combination of numerical options has been shown to be both accurate and robust in the context of resolving the disk buoyancy response to a low-mass planet \citep{ziampras-etal-2023b}.

\citet{ziampras-etal-2023b} showed that, even at high resolution and with a sophisticated numerical setup, \pluto{} struggles to capture the disk buoyancy response and the associated torque compared to finite-difference, upwind codes such as \fargo{} \citep{benitez-llambay-etal-2016}. For this purpose, while we carry out our radiative simulations with \pluto{}, we also perform a set of simulations in the adiabatic limit and with a simplified, local cooling prescription with both \pluto{} and \fargo{} \texttt{v.1.2}. In this way, we can gauge the extent to which \pluto{} might underestimate the disk's buoyancy response and correct for it, allowing us to estimate the planet's migration track in a scenario where the buoyancy torque is captured more appropriately.

Our \fargo{} setup is based on the hybrid MPI/OpenMP parallelism implemented by \citet{mcnally-etal-2019b} and solves the specific entropy $s$ instead of the energy equation in Eq.~\eqref{eq:navier-stokes-3}:
\begin{equation}
	\label{eq:navier-stokes-e}
	\DP{s}{t} + \vel\cdot\nabla s = Q^\prime,\qquad s\equiv\cv\log\left[\frac{T}{T_\text{ref}}\left(\frac{\rho}{\rho_\text{ref}}\right)^{1-\gamma}\right],
\end{equation}
where $T_\text{ref}$ and $\rho_\text{ref}$ are arbitrary constants. The thermal relaxation source term $Q^\prime$, when applicable, essentially represents the same physics as in Eq.~\eqref{eq:navier-stokes-3} but is rewritten in the specific entropy formulation.

In both codes our grid extends radially between 0.8--4\,au with a logarithmic spacing, vertically from the midplane up to $z=5H$ at $R_0=2\,$au, and spans the full azimuthal range 0--$2\pi$. We use a grid of $N_r\times N_\theta \times N_\phi = 528\times80\times2048$ cells, which achieves a resolution of 16 cells per scale height at $R_0$ in all directions. Similar to \citet{mcnally-etal-2020} and \citet{ziampras-etal-2023b}, we employ wave-damping zones near the radial boundaries of the domain following \citet{devalborro-etal-2006} and smooth the planet's potential with the cubic spline curve described in \citet{klahr-kley-2006}. The radiation energy is initialized as $E_\mathrm{rad,0} = \aR T_0^4$. At the radial and vertical edges of the domain, all parameters are set to their initial values. For $\Erad$, this implies that the disk is shielded from direct stellar irradiation, which is satisfied in our models (see green lines in Fig.~\ref{fig:disk-maps}).

The planet is held fixed at $R_0$ for 200 orbits and is then allowed to migrate for an additional 250, for a total of 450 orbits at $R_0$ by the end of the simulations. The planet grows over 200 orbits to its final mass of $\Mp=2\times10^{-5}\,\Mstar\approx6.7\,\mathrm{M}_\oplus$ using the formula by \citet{devalborro-etal-2006}. Since we do not consider the disk self-gravity, we subtract the azimuthal average of the gas surface density before computing the acceleration on the planet \citep{baruteau-masset-2008b}. Finally, we include the indirect term due to the star--planet system orbiting about their center of mass, and further include the indirect term due to star--disk interaction during the migration phase \citep{crida-etal-2022}. 

In our adiabatic runs, the source terms $Q$ and $Q^\prime$ in Eqs.~\eqref{eq:navier-stokes-3}~\&~\eqref{eq:navier-stokes-e} are set to zero. In our runs with a simplified local $\beta$ cooling prescription we implement a thermal relaxation term similar to \citet{gammie-2001} and \citet{mcnally-etal-2020} for \pluto{} and \fargo{}, respectively:
\begin{equation}
	\Qrelax = -\rho\cv\frac{T-T_0}{\btot}\OmegaK,\qquad\Qrelax^\prime = -\frac{s(\rho,T)-s(\rho,T_0)}{\btot}\OmegaK,
\end{equation}
where $\btot$ is given by Eqs.~\eqref{eq:beta}--\eqref{eq:btot}. Our fully radiative run in \pluto{} solves Eqs.~\eqref{eq:Erad}~\&~\eqref{eq:Qrad} instead. By comparing a fully radiative to a locally cooled model, we can assess the effect of radiation transport on the buoyancy torque as not only a cooling effect, but also a diffusion mechanism.

Finally, in order to isolate the effect of the buoyancy torque, we supplement our set of simulations with three vertically integrated $(R,\phi)$ models using \pluto{}. Similar to the 3D runs, we employ the following three cooling prescriptions:
\begin{itemize}
	\item \emph{adiabatic}: no cooling ($Q=0$),
	\item \emph{local cooling}: $\beta$ relaxation term following the prescription of surface and in-plane cooling in \citet{ziampras-etal-2024},
	\item \emph{radiative}: stellar irradiation, surface cooling, and in-plane radiative diffusion following the implementation of FLD in \citet{ziampras-etal-2020a}.
\end{itemize}
In these 2D models we use a Plummer potential for the planet's gravity, with a smoothing length $\epsilon=0.6\Hp$ \citep{mueller-kley-2012}.

\subsection{Static planet phase}

\begin{figure}
	\centering
	\includegraphics[width=\columnwidth]{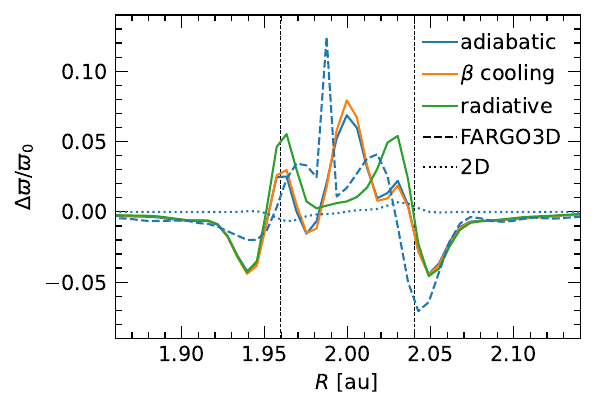}
	\caption{Azimuthally averaged perturbed vortensity for our models at $t=200$\,orbits. A vortensity excess inside the horseshoe region (marked by the dashed lines) is indicative of the dissipation of buoyancy oscillations in 3D runs.}
	\label{fig:1d-vortensity}
\end{figure}

During the first 200 orbits, the planet amasses a vortensity excess around the edge of its horseshoe region in all 3D models, with an additional excess in the center of the horseshoe region for the adiabatic and $\beta$-cooled models. We show the azimuthally averaged perturbed vortensity in Fig.~\ref{fig:1d-vortensity}. This already suggests that models with local cooling (i.e., that do not account for the effect of radiative diffusion) do not correctly capture the dissipation of buoyancy oscillations in the planet's horseshoe region, and therefore overestimate the associated vortensity generation.

\begin{figure}
	\centering
	\includegraphics[width=\columnwidth]{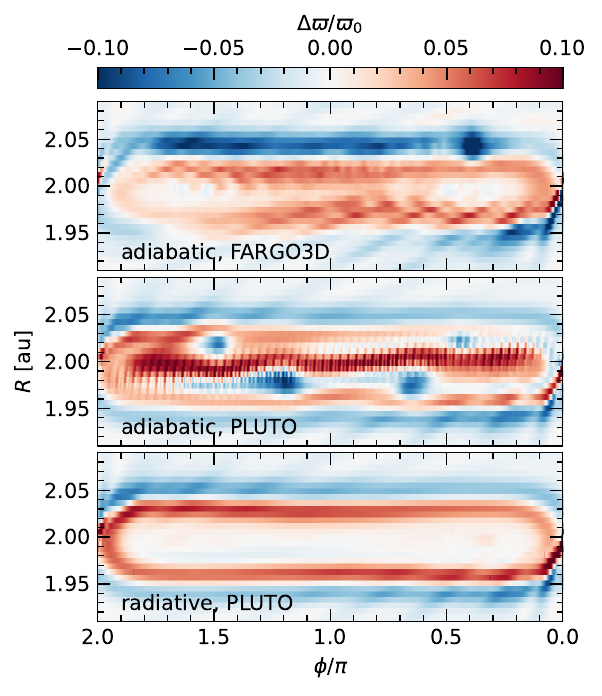}
	\caption{Perturbed vortensity for a selection of our 3D models at $t=200$~orbits. The vortensity excess in the adiabatic models is more centered to the planet's location, whereas the excess traces the edge of the horseshoe region in the radiative model. The planet is located at $\phi=0$ and moves towards the left.}
	\label{fig:2d-vortensity}
\end{figure}

We then plot perturbed vortensity heatmaps for the adiabatic and radiative 3D models in Fig.~\ref{fig:2d-vortensity}. Models with $\beta$ cooling are not shown as they are identical to their adiabatic counterparts. Similar to Fig.~\ref{fig:1d-vortensity}, we find that the vortensity excess is more centered to $R=\Rp$ in the adiabatic models, suggesting that higher-order buoyancy modes, which are excited radially closer to the planet, are damped due to radiative diffusion---fully consistent with the findings of \citet{yun-etal-2022}. The steeper vortensity gradients induced by the stronger vortensity growth in the adiabatic models also result in the excitation of vortices, which orbit around the horseshoe region and are not present in the radiative models. 

Overall, until the planet is allowed to migrate, our results are consistent with the findings of \citet{mcnally-etal-2020} and \citet{yun-etal-2022}, at least to an extent: the buoyancy response is partially damped in radiative models, although a vortensity excess in the horseshoe region is still visible. This indicates that the longer cooling timescales in the inner disk are indeed sufficient for buoyancy torques to operate, albeit not at the same level as in the adiabatic models. 

\subsection{Migrating planet phase}

\begin{figure}
	\centering
	\includegraphics[width=\columnwidth]{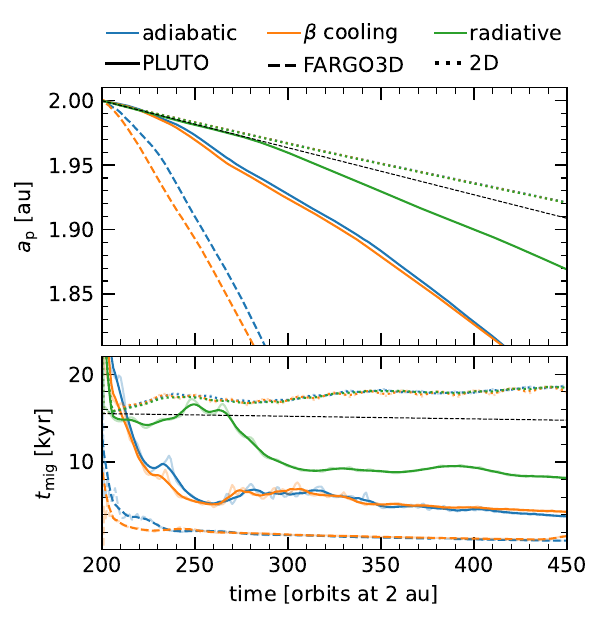}
	\caption{Migration tracks for all models. Top: semimajor axis of the planet as a function of time. Bottom: the migration timescale $\tau_\mathrm{mig}=\Rp/\dot{R}_\mathrm{p}$. Different colors represent different radiative prescriptions. Faint curves denote raw data, while solid curves denote a 20-orbit rolling average. Black dotted lines show the type-I migration track and cooling timescale. All 2D models overlap nearly perfectly.}
	\label{fig:tracks}
\end{figure}

We now allow the planet to migrate through the disk for an additional 250 orbits. The migration tracks for all models and the associated migration timescales $\tau_\mathrm{mig}=\Rp/\dot{R}_\mathrm{p}$ are shown in Fig.~\ref{fig:tracks}.
Several takeaways can be drawn from this figure:
\begin{itemize}
	\item In 2D, the planet migrates inward slower than the type-I rate, indicating a positive dynamical corotation torque (DCT) for all radiative treatments. This is expected given the very long cooling timescale and the lack of a buoyancy response in these 2D models, and leads to nearly perfectly overlapping tracks in all 2D models.
	\item The buoyancy response in our adiabatic 3D models induces a negative DCT, resulting in rapid inward migration. This is consistent with the findings of \citet{mcnally-etal-2020}.
	\item The planet initially migrates at the type-I rate in our radiative 3D models, but accelerates after $\approx80$~orbits. This suggests that the buoyancy-driven torque is operating, and reaches full strength as the planet continues to migrate. We investigate this behavior further below, but it is clear that the buoyancy torque is indeed active in our radiative models.
	\item Models with local cooling show a migration rate practically identical to the adiabatic models for both codes, indicating that the buoyancy torque is higher than in the radiative case and highlighting the need for realistic radiative modeling in order to correctly capture the buoyancy response of the disk.
\end{itemize}

\begin{figure}
	\centering
	\includegraphics[width=\columnwidth]{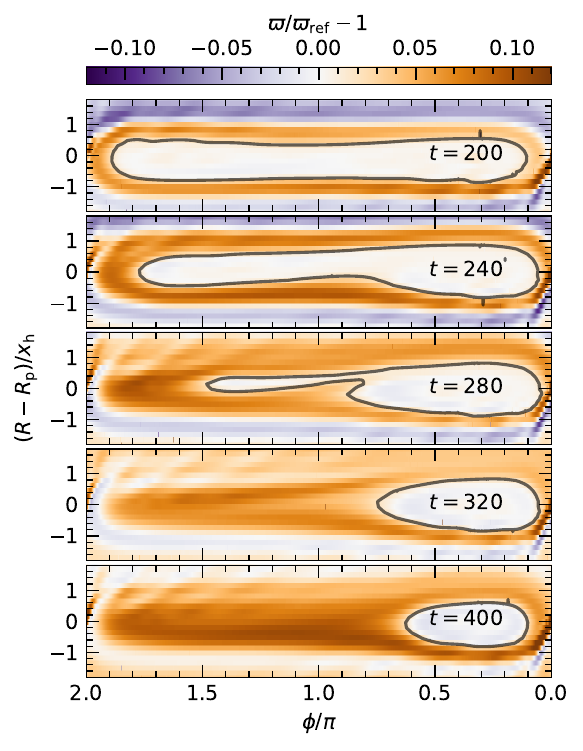}
	\caption{Vortensity heatmaps for the radiative 3D model at several snapshots during the planet's migration, highlighting the section of the corotating region where gas maintains its original vortensity. That section initially spans the full width of the horseshoe region, but shrinks to approximately a third of its initial size as the flow changes while the planet migrates. The planet then accelerates once this transition is complete.}
	\label{fig:rad-vortensity}
\end{figure}

To address the ``knee'' in the migration track of the 3D radiative model, we plot vortensity heatmaps near the planet's corotating region for this model and for several snapshots in Fig.~\ref{fig:rad-vortensity}. 
While the corotating region of the planet initially spans the full azimuthal range, it shrinks to a tadpole-shaped libration island \citep{masset-papaloizou-2003,papaloizou-etal-2007} ahead of the planet as the latter migrates inwards \citep[see also Fig.~18 in][]{mcnally-etal-2020}. At the same time, vortensity is continuously generated due to the dissipation of buoyancy modes close to the planet. Tracing the section of the corotating region where the gas maintains its original vortensity $\varpi_\mathrm{ref}=\varpi(R_0)$ shows that this section shrinks abruptly to approximately a third of its initial azimuthal extent after $\approx80$~orbits, at which point the planet accelerates inwards.

Observing the migration timescale in the bottom panel of Fig.~\ref{fig:tracks}, we find that this delayed transition to faster migration is present to an extent in all 3D models and is consistent with the libration timescale
\begin{equation}
	\label{eq:libration}
	\tau_\mathrm{lib} = \frac{8\pi \Rp}{3\Omega_\mathrm{p}\xh} \approx 72~\mathrm{orbits},
\end{equation}
which is the typical time it takes a gas streamline at the edge of the corotating region to complete a closed orbit and therefore allow the torque on the migrating planet to stabilize. We note that the planet's migration rate is roughly constant for all 3D models after $\approx80$ orbits, in line with this argument.

Consistent with the findings of \citet{ziampras-etal-2023b}, we find that \pluto{} underestimates the buoyancy torque compared to \fargo{}. In our models this translates to a factor of $\tau^\mathrm{Pluto}_\mathrm{mig}/\tau^\mathrm{Fargo}_\mathrm{mig}\approx3.5$ between the two codes for both adiabatic and $\beta$-cooled models. It is tempting to also scale the migration timescales of our radiative models by this factor, but we caution that the thermal diffusion associated with FLD results in some physical damping of the buoyancy response \citep[see][]{ziampras-etal-2023b}. As a result, the final migration timescale of the planet in a radiative disk model could in principle be up to 3.5$\times$ shorter that what we find in our \pluto{} models, but the difference between the two codes is likely much smaller in the radiative case.

Regardless, we stress that the buoyancy response exerts a substantial negative DCT on the planet in our radiative model, in contrast to the findings of \citet{yun-etal-2022} who concluded that the buoyancy torque is quenched in radiative disks. This does not invalidate their results for their disk model, but rather shows that whether the disk buoyancy response can exert a significant torque on a migrating planet is highly sensitive to the underlying disk conditions. In the next section we compare the excited buoyancy modes between the two models in an attempt to identify key differences in the buoyancy response of the disk.

\subsection{Buoyancy mode excitation}
\label{sub:buoyancy-mode-excitation}

In the previous section we showed that the disk buoyancy response can exert a significant torque on a migrating planet in favorable conditions, even when radiative cooling is considered. At the same time, \citet{yun-etal-2022} showed that the buoyancy torque is already quenched when $\bbuoy/\bdiff\lesssim0.1$ at $z\sim2H$. Here, we attempt to quantify the difference in the buoyancy response between the two disk models.

To that end, we carry out an additional set of short-term, global simulations with a fixed planet using \pluto{}, and compare the amplitude and phase of the buoyancy modes generated by the planet in the disk model presented in \citet{yun-etal-2022} and our hybrid model (see Sect.~\ref{sec:cooling-timescale-maps}). For these models the planet grows over one orbit and the simulation runtime is 10 orbits, similar to \citet{ziampras-etal-2023b}.

\begin{figure}
	\centering
	\includegraphics[width=\columnwidth]{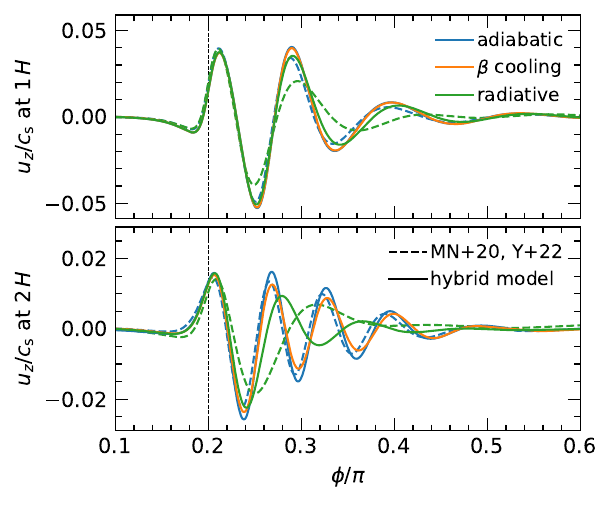}
	\caption{Azimuthal slice of the gas vertical velocity $u_z/\cs$ at $R=\Rp-\xh$ for the disk models of \citet{yun-etal-2022} and our hybrid model.  A vertical line marks the planet's azimuthal location. Buoyancy modes are more strongly damped in radiative models, and have a longer azimuthal wavelength for the model in \citet{yun-etal-2022}. }
	\label{fig:amplitude}
\end{figure}

Figure~\ref{fig:amplitude} shows an azimuthal slice of the gas vertical velocity $u_z$ normalized to the local sound speed at $z=\{H, 2H\}$ and $R=\Rp-\xh$, tracing the excited buoyancy oscillations at the inner edge of the horseshoe region. We find that:
\begin{itemize}
	\item Both models agree very well in the adiabatic case (blue curves), with our model showing very slightly stronger buoyancy modes on average. This difference is due to the disk flaring in our model, which results in a slightly smaller aspect ratio at that location. We find the opposite behavior at $R=\Rp+\xh$.
	\item Cooling damps buoyancy modes, more so for fully radiative (green) than for $\beta$ models (orange). This further highlights the need for realistic radiative modeling, and is consistent with the faster inward migration found for $\beta$ models in Fig.~\ref{fig:tracks}.
	\item Buoyancy modes in our radiative models (green) are more spaced out in azimuth compared to the adiabatic and $\beta$-cooled cases. This is consistent with the findings of \citet{yun-etal-2022}, and is most likely due to the diffusive nature of FLD.
\end{itemize}
The latter in particular might be a secondary reason for the weaker buoyancy torque, as a combination of weaker and fewer buoyancy modes will quickly diminish the rate at which vortensity can be generated in the corotating region. We expect that a ``cutoff'' in the buoyancy response of the disk will occur for $\bbuoy/\btot\gtrsim0.01$ at $z\sim2H$, but finding the exact value is beyond the scope of this work.

\section{Discussion}
\label{sec:discussion}

In this section we discuss the implications and limitations of our findings. We then outline the current challenges in planet migration modeling, and possible pathways to reconciling with the observed exoplanet population.

\subsection{Vortensity growth due to a radial entropy gradient}
\label{sub:entropy}

In this work we focused on the vortensity generation due to the disk buoyancy response. We note that, in principle, a radial entropy gradient can also lead to vortensity spikes around the edges of the horseshoe region \citep{masset-casoli-2009,paardekooper-etal-2010a}, although this has only been investigated in 2D. In our models, the entropy function is $P/\rho^\gamma\propto R^{0.48}$ and $P_\text{2D}/\Sigma^\gamma\propto R^{-0.03}$ in 3D and 2D, respectively. Thus, we can expect that a fraction of the vortensity growth shown for 3D models in Figs.~\ref{fig:1d-vortensity}~\&~\ref{fig:2d-vortensity} could be due to the contribution of this mechanism, rather than the disk buoyancy response \citep[see also Fig.~14 in][whose model satisfied $P/\rho^\gamma\propto R^{-0.4}$]{mcnally-etal-2020}.

Nevertheless, as found by \citet{mcnally-etal-2020}, the contribution by buoyancy modes---when they are active---dominates the vortensity growth in the horseshoe region, evident by the vortensity excess being centered on the planet in the adiabatic models while the planet is held fixed (see Fig.~\ref{fig:2d-vortensity}). Once the planet is allowed to migrate, the contribution by the buoyancy response dominates in the radiative model as well (see Fig.~\ref{fig:rad-vortensity}). Overall, we expect that the vortensity growth due to the disk buoyancy response is the dominant mechanism in accelerating migration in our models, and that the contribution by the radial entropy gradient is secondary.

\subsection{Observational constraints on the cooling timescale}
\label{sub:observational-evidence}

In Sects.~\ref{sec:cooling-timescale-maps}~\&~\ref{sec:results-simulations} we showed that, for the disk buoyancy response to drive meaningful vortensity generation, a very long cooling timescale (technically, a large ratio $\btot/\bbuoy$) is required. This condition is expected to be met in the inner, optically thick region of protoplanetary disks, but it is worth pointing out that $\btot$ will depend sensitively on the disk density and temperature as well as the properties of dust particles, which facilitate cooling.

Assuming a passive, irradiated disk model allows us to constrain its temperature structure \citep{chiang-goldreich-1997}, but the gas surface density and the fraction of small, efficiently-cooling dust grains are both largely unknown. With recent observational evidence\footnote[1]{Based on unpublished data from the AGE-PRO ALMA large program, private communication.} pointing towards relatively low disk masses of $\langle M_\mathrm{disk}\rangle \lesssim 0.01\,\Mstar$, and assuming a median disk cutoff radius of $\langle R_\mathrm{out}\rangle\sim30$\,au, a median stellar mass of $\langle\Mstar\rangle\approx0.5\,\Msun$, and $\Sigma\propto R^{-1}$, we find that the surface density at 1\,au for a typical star should be of the order of $\lesssim400$\,g/cm${}^2$. At the same time, dust growth models predict that a significant fraction of the submicron--micron-sized grains should be depleted within $\sim100$ local orbits as small grains coagulate to larger grains that no longer contribute to cooling, significantly lowering the opacity \citep[for a review, see][]{birnstiel-2023}.

As a result, we can expect that we could easily be overestimating the cooling timescale by one, two, or more orders of magnitude, and that buoyancy-driven torques should not be active for the typical planet-forming disk, or be limited to a very narrow radial range near the magnetically active inner rim of the disk \citep[see e.g.,][]{flock-etal-2017a}. 

\subsection{Additional physical mechanisms}
\label{sub:additional-physics}

While the previous paragraph suggests that buoyancy torques might not be ubiquitous, this does not imply that the stalling effect of the dynamical corotation torque (DCT) can efficiently slow down inward migration. \citet{ziampras-etal-2024} showed that for intermediate cooling timescales ($\btot\sim0.3$--100) the DCT is significantly weaker due to cooling-induced vortensity generation in the corotating region of a low-mass planet. As a result, a low-mass planet can be efficiently transported to the optically thick, slowly-cooling inner disk where $\btot\sim1000$ and buoyancy torques are active. The combined effect of these two mechanisms highlights the importance of radiation transport in hydrodynamical modeling and suggests that additional physical processes are necessary to prevent rapid inward migration and thus explain the population of super-Earths and sub-Neptunes at separations of $\lesssim2$\,au from their host stars.

Directly relevant to our discussion of dust coagulation in Sect.~\ref{sub:observational-evidence}, \citet{guilera-etal-2023} showed that a low-mass planet can efficiently trap a large amount of dust in its corotating region. The torque exerted by this concentration of solids was shown to be positive and comparable to or even larger in magnitude than the Lindblad torque for Stokes numbers of $\mathrm{St}\sim0.1$, able to stop or even reverse migration. Given that $\mathrm{St}\sim\Sigma^{-1}$, achieving $\mathrm{St}\sim0.1$ at 10\,au would require the presence of $\gtrsim$cm-sized grains, even with the lower surface densities discussed in Sect.~\ref{sub:observational-evidence}. This requirement is in tension with both observational evidence that point to a typical maximum grain size $a_\mathrm{max}\lesssim3$\,mm \citep{sierra-etal-2021}, as well as dust evolution models \citep[$a_\mathrm{max}\lesssim1$\,mm, see e.g.,][]{birnstiel-2023,dominik-dullemond-2024}. Nevertheless, dust dynamics could help slow down planet migration to an extent even for $\mathrm{St}\sim0.01$.

Moreover, thermal torques due to the planet's accretion luminosity could further slow down or reverse the direction of migration \citep{benitez-etal-2015,masset-2017}. Combining dust dynamics with an accretion luminosity due to a pebble flux, \citet{cummins-etal-2022} showed that the picture complicates significantly as the edge of the horseshoe region becomes unstable to the Rossby Wave Instability \citep[RWI,][]{lovelace-1999}, leading to vortices which can induce a stochastic element to the planet's migration \citep{lega-etal-2021}. We note, however, that none of these models included both a dust component and a self-consistent treatment of radiative cooling. Combining these mechanisms in one model while maintaining control over all processes involved is a challenge in itself, and will be the focus of follow-up work.

\subsection{Transition to type-II regime and the ``inertial limit''}

In the above, we used the term ``low-mass'' in a rather liberal way, referring in fact to planets well below the gap-opening or \emph{thermal} mass $\Mth\approx h^3\Mstar$ and below the disk feedback mass \citep{goodman-rafikov-2001,rafikov-2002}. Of course, in a cold, passively heated protoplanetary disk, where we typically expect $h\sim0.02$ at 1\,au, even an Earth-mass-sized planet can imprint on its disk and open a partial gap, slowing down migration. This concept of an ``inertial limit'' \citep{hourigan-ward-1984,ward-hourigan-1989,ward-1997a} could in theory prevent the rapid inward migration discussed above, although \citet{mcnally-etal-2019a} showed that intermittent vortex formation and dissipation can sustain an inward track. Based on the above paragraphs, investigating the effect of dust dynamics and vortex activity in disk models with realistic hydrodynamics is necessary in approaching a definitive answer to how low-mass planets actually interact with their surrounding disk.

\section{Summary}
\label{sec:summary}

We have investigated the presence of a negative dynamical corotation torque driven by the buoyancy response of a protoplanetary disk in the presence of a migrating planet \citep{zhu-etal-2012, lubow-zhu-2014, mcnally-etal-2020}. Our approach involved constructing a quasi-realistic yet computationally feasible disk model and then carrying out global, inviscid, radiation hydrodynamics simulations in the flux-limited diffusion (FLD) approximation with an embedded migrating planet.

We first computed the cooling timescale $\btot$ for a disk model used by \citet{mcnally-etal-2020} and \citet{yun-etal-2022}, and confirmed that the ratio $\bbuoy/\btot$ is $\lesssim0.01$ at $z\sim2H$ for $R\sim2$\,au, consistent with the quenching of buoyancy torques in their work. We then constructed a more realistic, passive, colder and flared disk model, and found that the conditions for the buoyancy torque to operate are more favorable at the same radius. However, due to the prohibitive computational cost of this model, we reach a compromise by constructing a hybrid setup that combines the reference parameters of the model used by \citet{yun-etal-2022} with the radial structure and longer cooling timescales of our realistic setup. We used this model in our hydrodynamical simulations.

Our numerical calculations showed that the planet migrates inwards significantly faster than in equivalent 2D models, indicating that the buoyancy torque is indeed active in the inner disk even when radiative diffusion is considered. In our radiative models, the planet's migration track exhibits a ``knee'' during its orbital evolution. Further analysis showed that this happens over a libration timescale, as the corotating region shrinks to a tadpole-shaped libration island and the torque on the planet stabilizes to a more negative value sustained by the vortensity generation due to the dissipation of buoyancy modes inside the planet's corotating region.

At the same time, models with a local, $\beta$ cooling prescription matched the migration rate of our adiabatic models for both \pluto{} and \fargo{}. Comparing individual buoyancy modes between our models revealed that FLD not only damps the disk buoyancy response, but also stretches the modes in azimuth, resulting in fewer and weaker oscillations and therefore much less efficient vortensity growth in the planet's corotating region \citep[consistent with][]{yun-etal-2022}. These findings highlight that the diffusive component of realistic radiation transport is necessary to correctly capture the buoyancy response of the disk, and further that the buoyancy response of the disk is likely to be rapidly quenched for larger $\bbuoy/\btot$, as weaker and fewer buoyancy modes will be excited.

Comparing \pluto{} with \fargo{}, we found that the former underestimates the buoyancy torque by a factor of $\approx3.5$, consistent with the findings of \citet{ziampras-etal-2023b}. While in the adiabatic and locally cooled models this discrepancy is of numerical origin, the thermal diffusion due to FLD should result in a physical damping of the buoyancy response in our radiative models. It is therefore plausible that the true migration timescale of the planet in a radiative model with \fargo{} could be up to 3.5$\times$ shorter than what we find in our \pluto{} models, although we expect the two codes to largely agree in the radiative case due to the damping associated with radiative diffusion.

Overall, our results show that the buoyancy torque can indeed reverse the dynamical corotation torque acting on a migrating planet in a realistic protoplanetary disk, and highlight the sensitivity of the buoyancy torque to the disk model and treatment of radiation transport. This suggests that planet migration in the low-turbulence dead zone might not ever slow down due to dynamical corotation torques \citep{paardekooper-2014}, and that low-mass planets \citep[i.e., below the feedback regime, see][]{hourigan-ward-1984,ward-hourigan-1989,ward-1997a} will likely migrate rapidly to the inner edge of the dead zone unless additional physical processes are considered. With that in mind, investigating the interplay of dust dynamics, radiation transport, and nonlinear processes such as gap opening and vortex formation will be critical to understanding the origin of super-Earths and sub-Neptunes in the disk dead zone.

\section*{Acknowledgements}
AZ would like to thank Will B{\'e}thune for their help in implementing and testing the FLD module, and Mario Flock for their suggestions and helpful discussions. This research utilized Queen Mary's Apocrita HPC facility, supported by QMUL Research-IT (http://doi.org/10.5281/zenodo.438045). This work was performed using the DiRAC Data Intensive service at Leicester, operated by the University of Leicester IT Services, which forms part of the STFC DiRAC HPC Facility (www.dirac.ac.uk). The equipment was funded by BEIS capital funding via STFC capital grants ST/K000373/1 and ST/R002363/1 and STFC DiRAC Operations grant ST/R001014/1. DiRAC is part of the National e-Infrastructure. AZ and RPN are supported by STFC grants ST/T000341/1 and ST/X000931/1. This project has received funding from the European Research Council (ERC) under the European Union's Horizon 2020 research and innovation programme (grant agreement No 101054502). All plots in this paper were made with the Python library \texttt{matplotlib} \citep{hunter-2007}.

\section*{Data Availability}

Data from our numerical models are available upon reasonable request to the corresponding author.

\bibliographystyle{mnras}
\bibliography{refs}

\begin{thebibliography}{}
\makeatletter
\relax
\def\mn@urlcharsother{\let\do\@makeother \do\$\do\&\do\#\do\^\do\_\do\%\do\~}
\def\mn@doi{\begingroup\mn@urlcharsother \@ifnextchar [ {\mn@doi@}
  {\mn@doi@[]}}
\def\mn@doi@[#1]#2{\def\@tempa{#1}\ifx\@tempa\@empty \href
  {http://dx.doi.org/#2} {doi:#2}\else \href {http://dx.doi.org/#2} {#1}\fi
  \endgroup}
\def\mn@eprint#1#2{\mn@eprint@#1:#2::\@nil}
\def\mn@eprint@arXiv#1{\href {http://arxiv.org/abs/#1} {{\tt arXiv:#1}}}
\def\mn@eprint@dblp#1{\href {http://dblp.uni-trier.de/rec/bibtex/#1.xml}
  {dblp:#1}}
\def\mn@eprint@#1:#2:#3:#4\@nil{\def\@tempa {#1}\def\@tempb {#2}\def\@tempc
  {#3}\ifx \@tempc \@empty \let \@tempc \@tempb \let \@tempb \@tempa \fi \ifx
  \@tempb \@empty \def\@tempb {arXiv}\fi \@ifundefined
  {mn@eprint@\@tempb}{\@tempb:\@tempc}{\expandafter \expandafter \csname
  mn@eprint@\@tempb\endcsname \expandafter{\@tempc}}}

\bibitem[\protect\citeauthoryear{{Bai} \& {Stone}}{{Bai} \&
  {Stone}}{2013}]{bai-stone-2013}
{Bai} X.-N.,  {Stone} J.~M.,  2013, \mn@doi [\apj]
  {10.1088/0004-637X/769/1/76}, \href
  {https://ui.adsabs.harvard.edu/abs/2013ApJ...769...76B} {769, 76}

\bibitem[\protect\citeauthoryear{Balay, Gropp, McInnes  \& Smith}{Balay
  et~al.}{1997}]{petsc-efficient}
Balay S.,  Gropp W.~D.,  McInnes L.~C.,   Smith B.~F.,  1997, in Arge E.,
  Bruaset A.~M.,   Langtangen H.~P.,  eds, Modern Software Tools in Scientific
  Computing. Birkh{\"{a}}user Press, pp 163--202

\bibitem[\protect\citeauthoryear{Balay et~al.,}{Balay
  et~al.}{2024a}]{petsc-web-page}
Balay S.,  et~al., 2024a, {PETS}c {W}eb page, \url{https://petsc.org/}, \url
  {https://petsc.org/}

\bibitem[\protect\citeauthoryear{Balay et~al.,}{Balay
  et~al.}{2024b}]{petsc-user-ref}
Balay S.,  et~al., 2024b, Technical Report ANL-21/39 - Revision 3.21,
  {PETSc/TAO} Users Manual.
Argonne National Laboratory, \mn@doi{10.2172/2205494}

\bibitem[\protect\citeauthoryear{{Baruteau} \& {Masset}}{{Baruteau} \&
  {Masset}}{2008}]{baruteau-masset-2008b}
{Baruteau} C.,  {Masset} F.,  2008, \mn@doi [\apj] {10.1086/529487}, \href
  {https://ui.adsabs.harvard.edu/abs/2008ApJ...678..483B} {678, 483}

\bibitem[\protect\citeauthoryear{{Ben{\'\i}tez-Llambay} \&
  {Masset}}{{Ben{\'\i}tez-Llambay} \&
  {Masset}}{2016}]{benitez-llambay-etal-2016}
{Ben{\'\i}tez-Llambay} P.,  {Masset} F.~S.,  2016, \mn@doi [\apjs]
  {10.3847/0067-0049/223/1/11}, \href
  {https://ui.adsabs.harvard.edu/abs/2016ApJS..223...11B} {223, 11}

\bibitem[\protect\citeauthoryear{{Ben{\'\i}tez-Llambay}, {Masset},
  {Koenigsberger}  \& {Szul{\'a}gyi}}{{Ben{\'\i}tez-Llambay}
  et~al.}{2015}]{benitez-etal-2015}
{Ben{\'\i}tez-Llambay} P.,  {Masset} F.,  {Koenigsberger} G.,   {Szul{\'a}gyi}
  J.,  2015, \mn@doi [\nat] {10.1038/nature14277}, \href
  {https://ui.adsabs.harvard.edu/abs/2015Natur.520...63B} {520, 63}

\bibitem[\protect\citeauthoryear{{Ben{\'\i}tez-Llambay}, {Krapp}  \&
  {Pessah}}{{Ben{\'\i}tez-Llambay} et~al.}{2019}]{benitez-llambay-etal-2019}
{Ben{\'\i}tez-Llambay} P.,  {Krapp} L.,   {Pessah} M.~E.,  2019, \mn@doi
  [\apjs] {10.3847/1538-4365/ab0a0e}, \href
  {https://ui.adsabs.harvard.edu/abs/2019ApJS..241...25B} {241, 25}

\bibitem[\protect\citeauthoryear{{Birnstiel}}{{Birnstiel}}{2023}]{birnstiel-2023}
{Birnstiel} T.,  2023, \mn@doi [arXiv e-prints] {10.48550/arXiv.2312.13287},
  \href {https://ui.adsabs.harvard.edu/abs/2023arXiv231213287B} {p.
  arXiv:2312.13287}

\bibitem[\protect\citeauthoryear{{Chiang} \& {Goldreich}}{{Chiang} \&
  {Goldreich}}{1997}]{chiang-goldreich-1997}
{Chiang} E.~I.,  {Goldreich} P.,  1997, \mn@doi [\apj] {10.1086/304869}, \href
  {https://ui.adsabs.harvard.edu/abs/1997ApJ...490..368C} {490, 368}

\bibitem[\protect\citeauthoryear{{Commer{\c{c}}on}, {Teyssier}, {Audit},
  {Hennebelle}  \& {Chabrier}}{{Commer{\c{c}}on}
  et~al.}{2011}]{commercon-etal-2011}
{Commer{\c{c}}on} B.,  {Teyssier} R.,  {Audit} E.,  {Hennebelle} P.,
  {Chabrier} G.,  2011, \mn@doi [\aap] {10.1051/0004-6361/201015880}, \href
  {https://ui.adsabs.harvard.edu/abs/2011A&A...529A..35C} {529, A35}

\bibitem[\protect\citeauthoryear{{Crida}, {Griveaud}, {Lega}, {Masset},
  {Morbidelli}, {Kloster}, {Marques}  \& {Minker}}{{Crida}
  et~al.}{2022}]{crida-etal-2022}
{Crida} A.,  {Griveaud} P.,  {Lega} E.,  {Masset} F.,  {Morbidelli} A.,
  {Kloster} D.,  {Marques} L.,   {Minker} K.,  2022, in {Richard} J.,  et~al.,
  eds, SF2A-2022: Proceedings of the Annual meeting of the French Society of
  Astronomy and Astrophysics. Eds.: J. Richard. pp 315--317

\bibitem[\protect\citeauthoryear{{Cummins}, {Owen}  \& {Booth}}{{Cummins}
  et~al.}{2022}]{cummins-etal-2022}
{Cummins} D.~P.,  {Owen} J.~E.,   {Booth} R.~A.,  2022, \mn@doi [\mnras]
  {10.1093/mnras/stac1819}, \href
  {https://ui.adsabs.harvard.edu/abs/2022MNRAS.515.1276C} {515, 1276}

\bibitem[\protect\citeauthoryear{{Dominik} \& {Dullemond}}{{Dominik} \&
  {Dullemond}}{2024}]{dominik-dullemond-2024}
{Dominik} C.,  {Dullemond} C.~P.,  2024, \mn@doi [\aap]
  {10.1051/0004-6361/202347716}, \href
  {https://ui.adsabs.harvard.edu/abs/2024A&A...682A.144D} {682, A144}

\bibitem[\protect\citeauthoryear{{Dullemond}, {Ziampras}, {Ostertag}  \&
  {Dominik}}{{Dullemond} et~al.}{2022}]{dullemond-etal-2022}
{Dullemond} C.~P.,  {Ziampras} A.,  {Ostertag} D.,   {Dominik} C.,  2022,
  \mn@doi [\aap] {10.1051/0004-6361/202244218}, \href
  {https://ui.adsabs.harvard.edu/abs/2022A&A...668A.105D} {668, A105}

\bibitem[\protect\citeauthoryear{{Flock}, {Fromang}, {Turner}  \&
  {Benisty}}{{Flock} et~al.}{2017a}]{flock-etal-2017a}
{Flock} M.,  {Fromang} S.,  {Turner} N.~J.,   {Benisty} M.,  2017a, \mn@doi
  [\apj] {10.3847/1538-4357/835/2/230}, \href
  {https://ui.adsabs.harvard.edu/abs/2017ApJ...835..230F} {835, 230}

\bibitem[\protect\citeauthoryear{{Flock}, {Nelson}, {Turner}, {Bertrang},
  {Carrasco-Gonz{\'a}lez}, {Henning}, {Lyra}  \& {Teague}}{{Flock}
  et~al.}{2017b}]{flock-etal-2017b}
{Flock} M.,  {Nelson} R.~P.,  {Turner} N.~J.,  {Bertrang} G. H.~M.,
  {Carrasco-Gonz{\'a}lez} C.,  {Henning} T.,  {Lyra} W.,   {Teague} R.,  2017b,
  \mn@doi [\apj] {10.3847/1538-4357/aa943f}, \href
  {https://ui.adsabs.harvard.edu/abs/2017ApJ...850..131F} {850, 131}

\bibitem[\protect\citeauthoryear{Gammie}{Gammie}{2001}]{gammie-2001}
Gammie C.~F.,  2001, \mn@doi [\apj] {10.1086/320631}, 553, 174

\bibitem[\protect\citeauthoryear{{Goldreich} \& {Tremaine}}{{Goldreich} \&
  {Tremaine}}{1979}]{goldreich-tremaine-1979}
{Goldreich} P.,  {Tremaine} S.,  1979, \mn@doi [\apj] {10.1086/157448}, \href
  {https://ui.adsabs.harvard.edu/abs/1979ApJ...233..857G} {233, 857}

\bibitem[\protect\citeauthoryear{{Goldreich} \& {Tremaine}}{{Goldreich} \&
  {Tremaine}}{1980}]{goldreich-tremaine-1980}
{Goldreich} P.,  {Tremaine} S.,  1980, \mn@doi [\apj] {10.1086/158356}, \href
  {https://ui.adsabs.harvard.edu/abs/1980ApJ...241..425G} {241, 425}

\bibitem[\protect\citeauthoryear{{Goodman} \& {Rafikov}}{{Goodman} \&
  {Rafikov}}{2001}]{goodman-rafikov-2001}
{Goodman} J.,  {Rafikov} R.~R.,  2001, \mn@doi [\apj] {10.1086/320572}, \href
  {https://ui.adsabs.harvard.edu/abs/2001ApJ...552..793G} {552, 793}

\bibitem[\protect\citeauthoryear{{Gressel}, {Turner}, {Nelson}  \&
  {McNally}}{{Gressel} et~al.}{2015}]{gressel-2015}
{Gressel} O.,  {Turner} N.~J.,  {Nelson} R.~P.,   {McNally} C.~P.,  2015,
  \mn@doi [\apj] {10.1088/0004-637X/801/2/84}, \href
  {https://ui.adsabs.harvard.edu/abs/2015ApJ...801...84G} {801, 84}

\bibitem[\protect\citeauthoryear{{Guilera}, {Benitez-Llambay}, {Miller
  Bertolami}  \& {Pessah}}{{Guilera} et~al.}{2023}]{guilera-etal-2023}
{Guilera} O.~M.,  {Benitez-Llambay} P.,  {Miller Bertolami} M.~M.,   {Pessah}
  M.~E.,  2023, \mn@doi [\apj] {10.3847/1538-4357/acd2cb}, \href
  {https://ui.adsabs.harvard.edu/abs/2023ApJ...953...97G} {953, 97}

\bibitem[\protect\citeauthoryear{{Harten}}{{Harten}}{1983}]{harten-1983}
{Harten} A.,  1983, \mn@doi [Journal of Computational Physics]
  {10.1016/0021-9991(83)90136-5}, \href
  {https://ui.adsabs.harvard.edu/abs/1983JCoPh..49..357H} {49, 357}

\bibitem[\protect\citeauthoryear{{Hayashi}}{{Hayashi}}{1981}]{hayashi-1981}
{Hayashi} C.,  1981, \mn@doi [Progress of Theoretical Physics Supplement]
  {10.1143/PTPS.70.35}, \href
  {https://ui.adsabs.harvard.edu/abs/1981PThPS..70...35H} {70, 35}

\bibitem[\protect\citeauthoryear{{Hourigan} \& {Ward}}{{Hourigan} \&
  {Ward}}{1984}]{hourigan-ward-1984}
{Hourigan} K.,  {Ward} W.~R.,  1984, \mn@doi [\icarus]
  {10.1016/0019-1035(84)90136-2}, \href
  {https://ui.adsabs.harvard.edu/abs/1984Icar...60...29H} {60, 29}

\bibitem[\protect\citeauthoryear{Hunter}{Hunter}{2007}]{hunter-2007}
Hunter J.~D.,  2007, Computing In Science \& Engineering, 9, 90

\bibitem[\protect\citeauthoryear{{Klahr} \& {Kley}}{{Klahr} \&
  {Kley}}{2006}]{klahr-kley-2006}
{Klahr} H.,  {Kley} W.,  2006, \mn@doi [\aap] {10.1051/0004-6361:20053238},
  \href {https://ui.adsabs.harvard.edu/abs/2006A&A...445..747K} {445, 747}

\bibitem[\protect\citeauthoryear{{Kley}}{{Kley}}{1989}]{kley-1989}
{Kley} W.,  1989, \aap, \href
  {https://ui.adsabs.harvard.edu/abs/1989A&A...208...98K} {208, 98}

\bibitem[\protect\citeauthoryear{{Kolb}, {Stute}, {Kley}  \& {Mignone}}{{Kolb}
  et~al.}{2013}]{kolb-etal-2013}
{Kolb} S.~M.,  {Stute} M.,  {Kley} W.,   {Mignone} A.,  2013, \mn@doi [\aap]
  {10.1051/0004-6361/201321499}, \href
  {https://ui.adsabs.harvard.edu/abs/2013A&A...559A..80K} {559, A80}

\bibitem[\protect\citeauthoryear{{Lega} et~al.,}{{Lega}
  et~al.}{2021}]{lega-etal-2021}
{Lega} E.,  et~al., 2021, \mn@doi [\aap] {10.1051/0004-6361/202039520}, \href
  {https://ui.adsabs.harvard.edu/abs/2021A&A...646A.166L} {646, A166}

\bibitem[\protect\citeauthoryear{{Lega}, {Morbidelli}, {Nelson}, {Ramos},
  {Crida}, {B{\'e}thune}  \& {Batygin}}{{Lega} et~al.}{2022}]{lega-etal-2022}
{Lega} E.,  {Morbidelli} A.,  {Nelson} R.~P.,  {Ramos} X.~S.,  {Crida} A.,
  {B{\'e}thune} W.,   {Batygin} K.,  2022, \mn@doi [\aap]
  {10.1051/0004-6361/202141675}, \href
  {https://ui.adsabs.harvard.edu/abs/2022A&A...658A..32L} {658, A32}

\bibitem[\protect\citeauthoryear{{Levermore} \& {Pomraning}}{{Levermore} \&
  {Pomraning}}{1981}]{levermore-pomraning-1981}
{Levermore} C.~D.,  {Pomraning} G.~C.,  1981, \mn@doi [\apj] {10.1086/159157},
  \href {https://ui.adsabs.harvard.edu/abs/1981ApJ...248..321L} {248, 321}

\bibitem[\protect\citeauthoryear{{Lin} \& {Papaloizou}}{{Lin} \&
  {Papaloizou}}{1985}]{lin-papaloizou-1985}
{Lin} D.~N.~C.,  {Papaloizou} J.,  1985, in {Black} D.~C.,  {Matthews} M.~S.,
  eds, Protostars and Planets II. pp 981--1072

\bibitem[\protect\citeauthoryear{{Lin} \& {Papaloizou}}{{Lin} \&
  {Papaloizou}}{1986}]{lin-papaloizou-1986}
{Lin} D.~N.~C.,  {Papaloizou} J.,  1986, \mn@doi [\apj] {10.1086/164653}, \href
  {https://ui.adsabs.harvard.edu/abs/1986ApJ...309..846L} {309, 846}

\bibitem[\protect\citeauthoryear{Lovelace, Li, Colgate  \& Nelson}{Lovelace
  et~al.}{1999}]{lovelace-1999}
Lovelace R. V.~E.,  Li H.,  Colgate S.~A.,   Nelson A.~F.,  1999, \mn@doi
  [\apj] {10.1086/306900}, 513, 805

\bibitem[\protect\citeauthoryear{{Lubow} \& {Zhu}}{{Lubow} \&
  {Zhu}}{2014}]{lubow-zhu-2014}
{Lubow} S.~H.,  {Zhu} Z.,  2014, \mn@doi [\apj] {10.1088/0004-637X/785/1/32},
  \href {https://ui.adsabs.harvard.edu/abs/2014ApJ...785...32L} {785, 32}

\bibitem[\protect\citeauthoryear{{Masset}}{{Masset}}{2000}]{masset-2000}
{Masset} F.,  2000, \mn@doi [\aaps] {10.1051/aas:2000116}, \href
  {http://adsabs.harvard.edu/abs/2000A%26AS..141..165M} {141, 165}

\bibitem[\protect\citeauthoryear{{Masset}}{{Masset}}{2017}]{masset-2017}
{Masset} F.~S.,  2017, \mn@doi [\mnras] {10.1093/mnras/stx2271}, \href
  {https://ui.adsabs.harvard.edu/abs/2017MNRAS.472.4204M} {472, 4204}

\bibitem[\protect\citeauthoryear{{Masset} \& {Ben{\'\i}tez-Llambay}}{{Masset}
  \& {Ben{\'\i}tez-Llambay}}{2016}]{masset-llambay-2016}
{Masset} F.~S.,  {Ben{\'\i}tez-Llambay} P.,  2016, \mn@doi [\apj]
  {10.3847/0004-637X/817/1/19}, \href
  {https://ui.adsabs.harvard.edu/abs/2016ApJ...817...19M} {817, 19}

\bibitem[\protect\citeauthoryear{{Masset} \& {Casoli}}{{Masset} \&
  {Casoli}}{2009}]{masset-casoli-2009}
{Masset} F.~S.,  {Casoli} J.,  2009, \mn@doi [\apj]
  {10.1088/0004-637X/703/1/857}, \href
  {https://ui.adsabs.harvard.edu/abs/2009ApJ...703..857M} {703, 857}

\bibitem[\protect\citeauthoryear{{Masset} \& {Papaloizou}}{{Masset} \&
  {Papaloizou}}{2003}]{masset-papaloizou-2003}
{Masset} F.~S.,  {Papaloizou} J.~C.~B.,  2003, \mn@doi [\apj] {10.1086/373892},
  \href {https://ui.adsabs.harvard.edu/abs/2003ApJ...588..494M} {588, 494}

\bibitem[\protect\citeauthoryear{{McNally}, {Nelson}, {Paardekooper}, {Gressel}
   \& {Lyra}}{{McNally} et~al.}{2017}]{mcnally-etal-2017}
{McNally} C.~P.,  {Nelson} R.~P.,  {Paardekooper} S.-J.,  {Gressel} O.,
  {Lyra} W.,  2017, \mn@doi [\mnras] {10.1093/mnras/stx2136}, \href
  {https://ui.adsabs.harvard.edu/abs/2017MNRAS.472.1565M} {472, 1565}

\bibitem[\protect\citeauthoryear{{McNally}, {Nelson}, {Paardekooper}  \&
  {Ben{\'\i}tez-Llambay}}{{McNally} et~al.}{2019a}]{mcnally-etal-2019a}
{McNally} C.~P.,  {Nelson} R.~P.,  {Paardekooper} S.-J.,
  {Ben{\'\i}tez-Llambay} P.,  2019a, \mn@doi [\mnras] {10.1093/mnras/stz023},
  \href {https://ui.adsabs.harvard.edu/abs/2019MNRAS.484..728M} {484, 728}

\bibitem[\protect\citeauthoryear{{McNally}, {Nelson}  \&
  {Paardekooper}}{{McNally} et~al.}{2019b}]{mcnally-etal-2019b}
{McNally} C.~P.,  {Nelson} R.~P.,   {Paardekooper} S.-J.,  2019b, \mn@doi
  [\mnras] {10.1093/mnrasl/slz118}, \href
  {https://ui.adsabs.harvard.edu/abs/2019MNRAS.489L..17M} {489, L17}

\bibitem[\protect\citeauthoryear{{McNally}, {Nelson}, {Paardekooper},
  {Ben{\'\i}tez-Llambay}  \& {Gressel}}{{McNally}
  et~al.}{2020}]{mcnally-etal-2020}
{McNally} C.~P.,  {Nelson} R.~P.,  {Paardekooper} S.-J.,
  {Ben{\'\i}tez-Llambay} P.,   {Gressel} O.,  2020, \mn@doi [\mnras]
  {10.1093/mnras/staa576}, \href
  {https://ui.adsabs.harvard.edu/abs/2020MNRAS.493.4382M} {493, 4382}

\bibitem[\protect\citeauthoryear{{Menou} \& {Goodman}}{{Menou} \&
  {Goodman}}{2004}]{menou-goodman-2004}
{Menou} K.,  {Goodman} J.,  2004, \mn@doi [\apj] {10.1086/382947}, \href
  {https://ui.adsabs.harvard.edu/abs/2004ApJ...606..520M} {606, 520}

\bibitem[\protect\citeauthoryear{{Mignone}, {Bodo}, {Massaglia}, {Matsakos},
  {Tesileanu}, {Zanni}  \& {Ferrari}}{{Mignone}
  et~al.}{2007}]{mignone-etal-2007}
{Mignone} A.,  {Bodo} G.,  {Massaglia} S.,  {Matsakos} T.,  {Tesileanu} O.,
  {Zanni} C.,   {Ferrari} A.,  2007, \mn@doi [The Astrophysical Journal
  Supplement Series] {10.1086/513316}, \href
  {https://ui.adsabs.harvard.edu/\#abs/2007ApJS..170..228M} {170, 228}

\bibitem[\protect\citeauthoryear{{Mignone}, {Flock}, {Stute}, {Kolb}  \&
  {Muscianisi}}{{Mignone} et~al.}{2012}]{mignone-etal-2012}
{Mignone} A.,  {Flock} M.,  {Stute} M.,  {Kolb} S.~M.,   {Muscianisi} G.,
  2012, \mn@doi [\aap] {10.1051/0004-6361/201219557}, \href
  {http://adsabs.harvard.edu/abs/2012A%26A...545A.152M} {545, A152}

\bibitem[\protect\citeauthoryear{{M{\"u}ller} \& {Kley}}{{M{\"u}ller} \&
  {Kley}}{2012}]{mueller-kley-2012}
{M{\"u}ller} T.~W.~A.,  {Kley} W.,  2012, \mn@doi [\aap]
  {10.1051/0004-6361/201118202}, \href
  {https://ui.adsabs.harvard.edu/abs/2012A&A...539A..18M} {539, A18}

\bibitem[\protect\citeauthoryear{{Nelson}, {Gressel}  \& {Umurhan}}{{Nelson}
  et~al.}{2013}]{nelson-etal-2013}
{Nelson} R.~P.,  {Gressel} O.,   {Umurhan} O.~M.,  2013, \mn@doi [\mnras]
  {10.1093/mnras/stt1475}, \href
  {https://ui.adsabs.harvard.edu/abs/2013MNRAS.435.2610N} {435, 2610}

\bibitem[\protect\citeauthoryear{{Ogilvie} \& {Lubow}}{{Ogilvie} \&
  {Lubow}}{2002}]{ogilvie-lubow-2002}
{Ogilvie} G.~I.,  {Lubow} S.~H.,  2002, \mn@doi [\mnras]
  {10.1046/j.1365-8711.2002.05148.x}, \href
  {https://ui.adsabs.harvard.edu/abs/2002MNRAS.330..950O} {330, 950}

\bibitem[\protect\citeauthoryear{{Paardekooper}}{{Paardekooper}}{2014}]{paardekooper-2014}
{Paardekooper} S.~J.,  2014, \mn@doi [\mnras] {10.1093/mnras/stu1542}, \href
  {https://ui.adsabs.harvard.edu/abs/2014MNRAS.444.2031P} {444, 2031}

\bibitem[\protect\citeauthoryear{{Paardekooper}, {Baruteau}, {Crida}  \&
  {Kley}}{{Paardekooper} et~al.}{2010a}]{paardekooper-etal-2010a}
{Paardekooper} S.~J.,  {Baruteau} C.,  {Crida} A.,   {Kley} W.,  2010a, \mn@doi
  [\mnras] {10.1111/j.1365-2966.2009.15782.x}, \href
  {https://ui.adsabs.harvard.edu/abs/2010MNRAS.401.1950P} {401, 1950}

\bibitem[\protect\citeauthoryear{{Paardekooper}, {Lesur}  \&
  {Papaloizou}}{{Paardekooper} et~al.}{2010b}]{paardekooper-etal-2010b}
{Paardekooper} S.-J.,  {Lesur} G.,   {Papaloizou} J. C.~B.,  2010b, \mn@doi
  [\apj] {10.1088/0004-637X/725/1/146}, \href
  {https://ui.adsabs.harvard.edu/abs/2010ApJ...725..146P} {725, 146}

\bibitem[\protect\citeauthoryear{{Paardekooper}, {Baruteau}  \&
  {Kley}}{{Paardekooper} et~al.}{2011}]{paardekooper-etal-2011}
{Paardekooper} S.~J.,  {Baruteau} C.,   {Kley} W.,  2011, \mn@doi [\mnras]
  {10.1111/j.1365-2966.2010.17442.x}, \href
  {https://ui.adsabs.harvard.edu/abs/2011MNRAS.410..293P} {410, 293}

\bibitem[\protect\citeauthoryear{{Paardekooper}, {Dong}, {Duffell}, {Fung},
  {Masset}, {Ogilvie}  \& {Tanaka}}{{Paardekooper}
  et~al.}{2022}]{paardekooper-etal-2022}
{Paardekooper} S.-J.,  {Dong} R.,  {Duffell} P.,  {Fung} J.,  {Masset} F.~S.,
  {Ogilvie} G.,   {Tanaka} H.,  2022, \mn@doi [arXiv e-prints]
  {10.48550/arXiv.2203.09595}, \href
  {https://ui.adsabs.harvard.edu/abs/2022arXiv220309595P} {p. arXiv:2203.09595}

\bibitem[\protect\citeauthoryear{{Papaloizou}, {Nelson}, {Kley}, {Masset}  \&
  {Artymowicz}}{{Papaloizou} et~al.}{2007}]{papaloizou-etal-2007}
{Papaloizou} J.~C.~B.,  {Nelson} R.~P.,  {Kley} W.,  {Masset} F.~S.,
  {Artymowicz} P.,  2007, in {Reipurth} B.,  {Jewitt} D.,   {Keil} K.,  eds,
  Protostars and Planets V. p.~655 (\mn@eprint {arXiv} {astro-ph/0603196}),
  \mn@doi{10.48550/arXiv.astro-ph/0603196}

\bibitem[\protect\citeauthoryear{{Pierens}}{{Pierens}}{2015}]{pierens-2015}
{Pierens} A.,  2015, \mn@doi [\mnras] {10.1093/mnras/stv2024}, \href
  {https://ui.adsabs.harvard.edu/abs/2015MNRAS.454.2003P} {454, 2003}

\bibitem[\protect\citeauthoryear{{Rafikov}}{{Rafikov}}{2002}]{rafikov-2002}
{Rafikov} R.~R.,  2002, \mn@doi [\apj] {10.1086/339399}, \href
  {https://ui.adsabs.harvard.edu/abs/2002ApJ...569..997R} {569, 997}

\bibitem[\protect\citeauthoryear{{Shakura} \& {Sunyaev}}{{Shakura} \&
  {Sunyaev}}{1973}]{shakura-sunyaev-1973}
{Shakura} N.~I.,  {Sunyaev} R.~A.,  1973, \aap, \href
  {https://ui.adsabs.harvard.edu/\#abs/1973A&A....24..337S} {500, 33}

\bibitem[\protect\citeauthoryear{{Sierra} et~al.,}{{Sierra}
  et~al.}{2021}]{sierra-etal-2021}
{Sierra} A.,  et~al., 2021, \mn@doi [\apjs] {10.3847/1538-4365/ac1431}, \href
  {https://ui.adsabs.harvard.edu/abs/2021ApJS..257...14S} {257, 14}

\bibitem[\protect\citeauthoryear{{Trac} \& {Pen}}{{Trac} \&
  {Pen}}{2004}]{trac-pen-2004}
{Trac} H.,  {Pen} U.-L.,  2004, \mn@doi [\na] {10.1016/j.newast.2004.02.002},
  \href {https://ui.adsabs.harvard.edu/abs/2004NewA....9..443T} {9, 443}

\bibitem[\protect\citeauthoryear{Van~Leer}{Van~Leer}{1974}]{vanleer-1974}
Van~Leer B.,  1974, Journal of computational physics, 14, 361

\bibitem[\protect\citeauthoryear{{Wang}, {Bai}  \& {Goodman}}{{Wang}
  et~al.}{2019}]{wang-etal-2019}
{Wang} L.,  {Bai} X.-N.,   {Goodman} J.,  2019, \mn@doi [\apj]
  {10.3847/1538-4357/ab06fd}, \href
  {https://ui.adsabs.harvard.edu/abs/2019ApJ...874...90W} {874, 90}

\bibitem[\protect\citeauthoryear{{Ward}}{{Ward}}{1997}]{ward-1997a}
{Ward} W.~R.,  1997, \mn@doi [\icarus] {10.1006/icar.1996.5647}, \href
  {https://ui.adsabs.harvard.edu/abs/1997Icar..126..261W} {126, 261}

\bibitem[\protect\citeauthoryear{{Ward} \& {Hourigan}}{{Ward} \&
  {Hourigan}}{1989}]{ward-hourigan-1989}
{Ward} W.~R.,  {Hourigan} K.,  1989, \mn@doi [\apj] {10.1086/168138}, \href
  {https://ui.adsabs.harvard.edu/abs/1989ApJ...347..490W} {347, 490}

\bibitem[\protect\citeauthoryear{{Yamaleev} \& {Carpenter}}{{Yamaleev} \&
  {Carpenter}}{2009}]{yamaleev-carpenter-2009}
{Yamaleev} N.~K.,  {Carpenter} M.~H.,  2009, \mn@doi [Journal of Computational
  Physics] {10.1016/j.jcp.2009.01.011}, \href
  {https://ui.adsabs.harvard.edu/abs/2009JCoPh.228.3025Y} {228, 3025}

\bibitem[\protect\citeauthoryear{{Yun}, {Kim}, {Bae}  \& {Han}}{{Yun}
  et~al.}{2022}]{yun-etal-2022}
{Yun} H.-G.,  {Kim} W.-T.,  {Bae} J.,   {Han} C.,  2022, \mn@doi [\apj]
  {10.3847/1538-4357/ac9185}, \href
  {https://ui.adsabs.harvard.edu/abs/2022ApJ...938..102Y} {938, 102}

\bibitem[\protect\citeauthoryear{{Zhu}, {Stone}  \& {Rafikov}}{{Zhu}
  et~al.}{2012}]{zhu-etal-2012}
{Zhu} Z.,  {Stone} J.~M.,   {Rafikov} R.~R.,  2012, \mn@doi [\apjl]
  {10.1088/2041-8205/758/2/L42}, \href
  {https://ui.adsabs.harvard.edu/abs/2012ApJ...758L..42Z} {758, L42}

\bibitem[\protect\citeauthoryear{{Ziampras}, {Ataiee}, {Kley}, {Dullemond}  \&
  {Baruteau}}{{Ziampras} et~al.}{2020}]{ziampras-etal-2020a}
{Ziampras} A.,  {Ataiee} S.,  {Kley} W.,  {Dullemond} C.~P.,   {Baruteau} C.,
  2020, \mn@doi [\aap] {10.1051/0004-6361/201936495}, \href
  {https://ui.adsabs.harvard.edu/abs/2020A&A...633A..29Z} {633, A29}

\bibitem[\protect\citeauthoryear{{Ziampras}, {Paardekooper}  \&
  {Nelson}}{{Ziampras} et~al.}{2023}]{ziampras-etal-2023b}
{Ziampras} A.,  {Paardekooper} S.-J.,   {Nelson} R.~P.,  2023, \mn@doi [\mnras]
  {10.1093/mnras/stad2692}, \href
  {https://ui.adsabs.harvard.edu/abs/2023MNRAS.525.5893Z} {525, 5893}

\bibitem[\protect\citeauthoryear{{Ziampras}, {Nelson}  \&
  {Paardekooper}}{{Ziampras} et~al.}{2024}]{ziampras-etal-2024}
{Ziampras} A.,  {Nelson} R.~P.,   {Paardekooper} S.-J.,  2024, \mn@doi [arXiv
  e-prints] {10.48550/arXiv.2402.00125}, \href
  {https://ui.adsabs.harvard.edu/abs/2024arXiv240200125Z} {p. arXiv:2402.00125}

\bibitem[\protect\citeauthoryear{{de Val-Borro} et~al.,}{{de Val-Borro}
  et~al.}{2006}]{devalborro-etal-2006}
{de Val-Borro} M.,  et~al., 2006, \mn@doi [\mnras]
  {10.1111/j.1365-2966.2006.10488.x}, \href
  {http://adsabs.harvard.edu/abs/2006MNRAS.370..529D} {370, 529}

\makeatother
\end{thebibliography}

\clearpage

\appendix

\section{Implementation of FLD solver}
\label{apdx:FLD-test}

Our FLD module closely follows the numerical implementation of \citet{kolb-etal-2013}. We nevertheless provide a brief description of our implementation, and carry out a supplementary validation test to ensure that it is correct.

\subsection{Numerics}

Similarly to the implementation of \citet{kolb-etal-2013} (see Sect.~3.2 therein), we write Eq.~\eqref{eq:Erad} in implicit form for $\Erad$, taking into account the finite-volume nature of \pluto{}. Following \citet{commercon-etal-2011}, the nonlinear term $\left(T^{n+1}\right)^4$ in $\Qrad$ is approximated as
\begin{equation}
	\label{eq:nonlinear}
	\left(T^{n+1}\right)^4 \approx 4\left(T^n\right)^3T^{n+1}-3\left(T^n\right)^4.
\end{equation}

We then solve the resulting linear system of equations using the BiCGStab method, implemented in the sparse matrix solver module of the \textsc{PetSc} library \citep{petsc-efficient,petsc-web-page,petsc-user-ref}. After obtaining the updated $\Erad$, we update the gas temperature as
\begin{equation}
	\label{eq:temperature}
	T^{n+1} = \frac{\kappaP c\left(3\aR T^4 + \Erad^{n+1}\right)\Delta t + \cv T}{\cv + 4\kappaP c \aR T^3 \Delta t}. 
\end{equation}

\subsection{Test problem: radiatively damped linear wave}

To test our implementation, we design a 1D test problem similar to \citet{benitez-llambay-etal-2019} where we excite small perturbations that correspond to eigenvectors of the coupled system of equations \eqref{eq:navier-stokes},~\eqref{eq:Erad}~\&~\eqref{eq:Qrad} and measure their oscillation frequency and damping rate. This both ensures that our implementation of the FLD solver for $\Erad$ in Eq.~\eqref{eq:Erad} is correct, and that the thermal energy and radiation energy field are coupled correctly in Eq.~\eqref{eq:Qrad}.

\subsubsection{Dispersion relation and eigenvectors}

We assume that $\lambda$, $\kappaP$, and $\kappaR$ are constant, and that all quantities $q \in \{\rho, u, P, \Erad\}$ can be expressed in the form $q=q_0 + q_1$, with $q_0$ being a constant background state and $q_1 = A q^\prime e^{ikx -\omega t}$ a wave-like perturbation with $A\ll 1$. By plugging them into the set of Eqs.~\eqref{eq:navier-stokes},~\eqref{eq:Erad}~\&~\eqref{eq:Qrad} and discarding terms of $\mathcal{O}(A^2)$ or higher we obtain
\begin{subequations}
	\label{eq:linearized}
	\begin{align}
		\label{eq:linearized-1}
		\DP{\rho_1}{t} + \rho_0 \DP{u_1}{x} = 0,
	\end{align}
	\begin{align}
		\label{eq:linearized-2}
		\DP{u_1}{t} = -\frac{1}{\rho_0}\DP{P_1}{x},
	\end{align}
	\begin{align}
		\label{eq:linearized-3}
		\DP{P_1}{t} = -\gamma P_0 \DP{u_1}{x} + (\gamma-1) \Qrad^\prime
	\end{align}
	\begin{align}
		\label{eq:linearized-4}
		\DP{{\Erad}_1}{t} - \frac{\lambda c}{\kappaR\rho_0}\DPP{{\Erad}_1}{x} = -\Qrad^\prime,
	\end{align}
\end{subequations}
where $\Qrad^\prime$ is the linearized radiation source term after expanding $T^4\propto P^4/\rho^4$ to first order and using that ${\Erad}_0 = \aR T_0^4$:
\begin{equation}
	\label{eq:Qrad-linearized}
	\Qrad^\prime = -\kappaP\rho_0c\left(4\frac{P_1}{P_0} - 4\frac{\rho_1}{\rho_0} - \frac{{\Erad}_1}{{\Erad}_0}\right).
\end{equation}

By combining Eqs.~\eqref{eq:linearized-1}--\eqref{eq:Qrad-linearized} and seeking a solution for $\omega$ we obtain the dispersion relation
\begin{equation}
	\label{eq:dispersion}
	\begin{split}
		\omega^4 &\left[\chi\omegaR\right] +\omega^3 \left[-\chi\omegac^2-(4+\chi)\omegaP\omegaR\right] +\omega^2 \left[\chi\omegaP\omegac^2+4\omegaP^2\right] \\+\omega&\left[-\gamma\chi\omegac^2\omegas^2-(4+\gamma\chi)\omegaP\omegaR\omegas^2\right]+4\omegaP\omegac^2\omegas^2 = 0,
	\end{split}
\end{equation}
where we have defined for convenience $\omegas = k {\cs}_0$, $\omegaR=\kappaR \rho_0 c$, $\omegaP=\kappaP \rho_0 c$, $\omegac=\sqrt{\lambda} k c$, and $\chi=e_0/{\Erad}_0$. This equation can be solved numerically to obtain the oscillation frequency and damping rate of radiatively damped sound waves, provided that $\text{Im}(\omega)\neq 0$. The components of the related eigenvectors are then 
\begin{equation}
	\label{eq:linearized2}
	\begin{split}
		\frac{u^\prime}{{\cs}_0} &= -\frac{\omega}{\omegas} \frac{\rho^\prime}{\rho_0},\\
		\frac{P^\prime}{P_0} &= -\frac{\omega^2}{\omegas^2} \frac{\rho^\prime}{\rho_0},\\
		\frac{\Erad^\prime}{{\Erad}_0} &= -\left(\gamma\chi\frac{\omega}{\omegaP} - 4\frac{\omega^2}{\omegas^2} - 4 + \chi\frac{\omega^3}{\omegaP\omegas^2}\right)\frac{\rho^\prime}{\rho_0}.\\
	\end{split}
\end{equation}

\subsubsection{Numerical solution}

\begin{figure}
	\centering
	\includegraphics[width=\columnwidth]{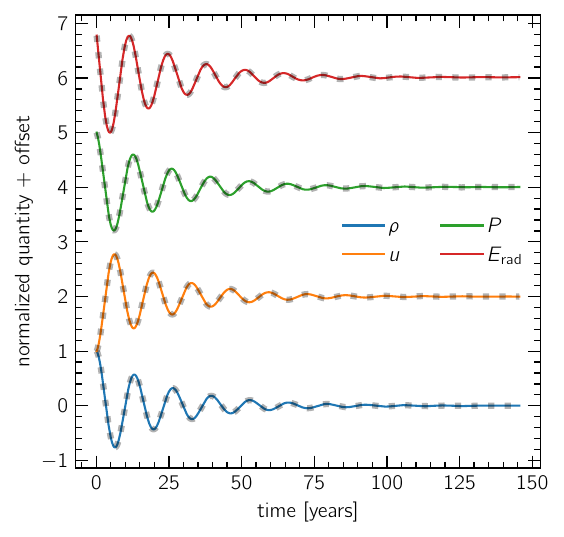}
	\caption{An example of fitted oscillations for our model with $\rho_0=10^{-12}$\,g/cm${}^3$ and $\kappa=1$\,cm${}^2$/g at $x=0$. All quantities have been normalized and shifted for clarity. Gray dots denote the analytical solutions.}
	\label{fig:dynamic-osc}
\end{figure}

In our numerical setup we consider a periodic Cartesian domain $x\in[0, L]$ with $L=1$\,au and 1000 uniformly spaced grid cells, and define $k=2\pi/L$. We set $\mu=2.353$, $\gamma=7/5$, $\lambda=1/3$, and $T_0=30$\,K. The opacity $\kappaR=\kappaP=\kappa$ is set to a constant value within each setup and varies between $0.003$--300\,$\text{cm}^2/\text{g}_\text{gas}$ throughout the set of models we run. The background state is given by
\begin{equation}
	\begin{split}
		\rho_0 &= \{10^{-13}, 10^{-12}\}~\text{g}/\text{cm}^3,&\qquad &u_0 = 0,\\
		P_0 &= \frac{\Rgas}{\mu} \rho_0 T_0,&\qquad&{\Erad}_0 = \aR T_0^4,
	\end{split}
\end{equation}
and the perturbed quantities are initialized as
\begin{equation}
	\begin{split}
		q_1 = A\left[\text{Re}\left(q^\prime\right)\cos(kx) - \text{Im}\left(q^\prime\right)\sin(kx)\right],
	\end{split}
\end{equation}
where $A=10^{-4}$ is the amplitude of the perturbation. We then integrate for a time corresponding to 150 years. Finally, we extract the oscillation frequency $\omega_\text{osc}$ and damping rate $\omega_\text{damp}$ from the perturbed quantities by fitting the gas pressure at $x=0$ as a function of time with a function of the form
\begin{equation}
	P(t) = P_0 + c_0  P_0 \sin(\omega_\text{osc} t + c_1) e^{\omega_\text{damp} t},
\end{equation}
where $c_0$ and $c_1$ are additional, free fitting parameters to account for the amplitude and phase of the excited eigenmode. An example of the fitted oscillation is shown in Fig.~\ref{fig:dynamic-osc} for all quantities.

\begin{figure}
	\centering
	\includegraphics[width=\columnwidth]{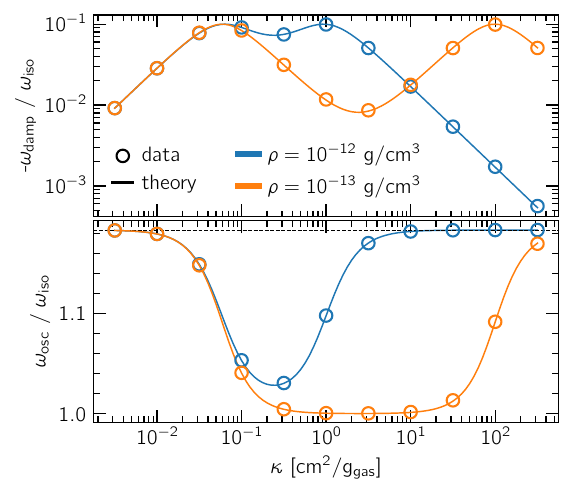}
	\caption{Results from the linear perturbation test, showing the damping $\omega_\mathrm{damp}$ (top) and oscillation frequency $\omega_\mathrm{osc}$ (bottom) as a function of the opacity $\kappa$ for two reference values of the background density $\rho_0$. We normalize $\omega$ to $\omega_\text{iso}=k {\csiso}_0$. The oscillation frequency fluctuates between the isothermal ($\omega_\text{iso}$) and adiabatic expectation ($\sqrt{\gamma}\omega_\text{iso}$, dashed black line), depending on the coupling between the thermal and radiation energy fields.}
	\label{fig:dynamic}
\end{figure}

We then present our results in Fig.~\ref{fig:dynamic}, showing that we recover the expected damping and oscillation frequencies to very good degree for all values of $\kappa$ and both reference values of $\rho_0$. We therefore conclude that our FLD module is suitable for use in our models.

\bsp	
\label{lastpage}
\end{document}